\documentclass[ twocolumn, trackchanges]{aastex701}
\accepted{March 26, 2026}

\defcitealias{Ferraro2009}{F09}
\defcitealias{Lugger2007}{L07}
\usepackage{graphicx}
\usepackage{subcaption}
\usepackage{float}
\usepackage{placeins}

\begin{document}

\title{Revisiting the Evidence for Double Sequences of Blue Straggler Stars in Globular Clusters}
\shorttitle{Are Double BSS Sequences Real?}
\shortauthors{G. Kumawat et al.}

\correspondingauthor{Gourav Kumawat}

\author[0009-0001-2894-0155]{Gourav Kumawat}
\affiliation{Department of Physics, University of Alberta, Edmonton AB T6G 2G7, Canada}
\email[show]{kumawat1@ualberta.ca}

\author[0000-0003-3944-6109]{Craig O. Heinke}
\affiliation{Department of Physics, University of Alberta, Edmonton AB T6G 2G7, Canada}
\email{heinke@ualberta.ca}

\author[0000-0003-3551-5090]{Alison Sills}
\affiliation{Department of Physics \& Astronomy, McMaster University, 1280 Main Street West, Hamilton, ON L8S 4M1, Canada}
\email{asills@mcmaster.ca}

\author[0000-0002-9673-7802]{Haldan N. Cohn}
\affiliation{Department of Astronomy, Indiana University, 727 E. Third St., Bloomington, IN 47405, USA}
\email{cohn@iu.edu}

\author[0000-0001-9953-6291]{Phyllis M. Lugger}
\affiliation{Department of Astronomy, Indiana University, 727 E. Third St., Bloomington, IN 47405, USA}
\email{lugger@iu.edu}

\author[0000-0002-1116-2553]{Christian Knigge}
\affiliation{Department of Physics and Astronomy. University of Southampton, Southampton, SO17 1BJ, UK}
\email{c.knigge@soton.ac.uk}

\author[0000-0002-3704-7190]{Andrea Dieball}
\affiliation{Helmholtz Institut für Strahlen und Kernphysik, Universität Bonn, Nussallee 14-16, 53115, Bonn, Germany}
\email{adieball@astro.uni-bonn.de}

\author[0009-0000-9304-9580]{Tyler Heise}
\affiliation{Department of Physics, University of Alberta, Edmonton AB T6G 2G7, Canada}
\email{trheise@ualberta.ca}

\begin{abstract}

Blue straggler stars (BSSs) are believed to form through  mass transfer in binary systems or  stellar collisions. The reported presence of double BSS sequences in some globular clusters (GCs) has been interpreted as evidence that these two formation channels produce distinct sequences in color--magnitude diagram (CMD). We reassess this claim using \textit{HST} UV Globular Cluster Survey (HUGS) photometry of 56 Galactic GCs. We used the Hartigan Dip Test to test bimodality, and Akaike model comparison to test whether BSS distance distributions are better described by a mixture of two unskewed Gaussians or a skewed unimodal Gaussian model. We find no strong statistical evidence for bimodality; no cluster yields a dip test $p$-value below 0.15, and Akaike model comparison favors the skewed unimodal model in 94 out of 112 cases. We re-examine NGC~7099 (M30), the prototypical case of a double BSS sequence, using three reductions of \textit{HST} data. We find bimodality is detected at p = $4 \times 10^{-3}$, vs. the originally reported p $\sim  10^{-5}$, in the original photometry. The observed uncertainties derived from the subgiant branch widths are comparable to the suggested separation between the proposed BSS sequences, making the detection of statistically significant bimodality challenging. Our results suggest that the dip between two BSS sequences in M30 photometry is a coincidence, and that later bifurcation claims can be explained as skew in the BSS color distribution, rather than two separate distributions.
\end{abstract}


\section{INTRODUCTION} \label{sec:intro}

Blue straggler stars (BSSs) were first identified by \citet{Sandage1953} in the globular cluster M3, where he noticed a distinct sequence of stars lying above and to the left of the main-sequence turnoff (MSTO) in the optical color–magnitude diagram (CMD). Their position in the CMD indicates that they are both bluer and more luminous than normal cluster main-sequence stars, implying higher masses \citep{Shara1997, Gilliland1998, Fiorentino2014}. However, in stellar systems as old as globular clusters (GCs), stars of such masses should have already evolved off the main sequence, suggesting that BSSs have undergone a rejuvenation process that increases their mass \citep{Bailyn1995}. Because of their higher masses relative to the average cluster stars, BSS progenitors are subject to mass segregation and dynamical friction, leading to their preferential concentration toward the dense cluster core \citep{Mapelli2004, Ferraro2012}. Consequently, the distribution and physical characteristics of BSSs provide valuable insights into both the stellar and dynamical evolution of their host clusters \citep{Ferraro2012}.

The origin of BSSs has been a subject of extensive study since their discovery. Two main formation pathways are generally accepted \citep[see][]{Mathieu25}: (1) mass transfer (MT) from a binary companion, ; and (2) direct stellar collisions, likely during a binary encounter. In the MT scenario, originally proposed by \citet{McCrea1964}, a (typically evolved) primary star fills its Roche lobe and transfers mass to a less massive secondary, rejuvenating it into a hotter, more massive main-sequence star. For an evolved companion, MT results in a binary of a BSS and a white dwarf, while a main-sequence companion leads to a (slow) stellar merger.  The slow merger phase goes through a long-lived contact binary (or W UMa, a peanut-shaped star; \citealt{Rucinski00}) phase before final merger \citep[e.g.][]{NelsonEggleton01}. Both the W UMa binary itself, and the final merger product, can appear as a BSS. The second formation channel involves direct collisions between stars \citep{Hills1976}. Although direct collisions of single non-giant stars are rare even in globular clusters, the presence of binary stars increases the likelihood of collisions, with numerical simulations demonstrating that such interactions are likely during binary–single or binary–binary encounters \citep{Lombardi1995, Fregeau2004}. 

In the cores of the densest GCs, where stellar interactions are frequent, both the collisional and MT mechanisms may operate simultaneously, contributing to the observed BSS population \citep{FusiPecci1992, Bailyn1992, Ferraro1995}. During the 1990s, stellar collisions were widely regarded as the dominant BSS formation channel in high-density environments, while MT in binary systems was considered more relevant in the low-density outskirts of clusters. However, \citet{Piotto2004} showed an anti-correlation between the BSS relative frequency and the total luminosity of the host cluster, with no statistically significant correlation between BSS frequency and the predicted collision rate. Instead, the BSS relative frequency seemed to clearly anti-correlate with cluster mass. This result suggested that BSS formation is not dominated by collisional processes, even in the densest clusters. By the late 2000s, accurate measurements of binary fractions and new theoretical work  indicated that MT is likely to dominate the production of BSSs even within dense cluster cores \citep{Sollima2008, Knigge2009, Milone12}. \citet{Ferraro2026} reported that the number of BSSs, when normalized to the sampled luminosity, correlates inversely with the central density of the parent cluster and positively with the binary fraction, implying that BSSs are largely produced by binaries in globular clusters.

\citet{Ferraro2009}, (hereafter \citetalias{Ferraro2009}) identified a double BSS sequence in the \textit{HST/WFPC2} $F555W-F814W$ CMD of the post-core-collapse (PCC) cluster NGC~7099 (M30). The two sequences are separated in color, with the blue BSS sequence suggested to match collisional models \citep{Sills2009} and the red sequence suggested to match models of mass-transfer (MT) binaries \citep{Tian2006}. This discovery suggested that both formation channels can operate simultaneously within dense cluster cores. The coexistence of the two populations was interpreted as a possible outcome of the core-collapse (CC) process, which enhances the stellar collision rate and may drive binaries toward the mass-transfer regime through dynamical interactions in the collapsing core. Given the evolutionary timescales of BSSs, the observation 
of two distinct sequences suggests that a CC event in NGC~7099 occurred within the last 1--2~Gyr (F09). 

Following this finding, bifurcations in the BSS population have been reported in several other clusters, including NGC~362 \citep{Dalessandro2013}, NGC~1261 \citep{Simunovic2014}, NGC~7078 (M15) \citep{Beccari2019}, NGC~6752 \citep{Cohn2021}, and NGC~6256 \citep{Cadelano2022}. A bifurcation has also been reported in the young Large Magellanic Cloud cluster NGC~2173 \citep{Li2018a}, though its reality has been questioned \citep{Dalessandro2019a,Dalessandro2019b}. 
If NGC 2173 indeed has a BSS bifurcation, both BSS sequences would have to be produced by primordial binaries, since it is too young and low-density to have experienced substantial collisions.

Several theoretical works have attempted to explain one or both BSS sequences. \citet{Xin15} computed mass-transfer BSS products that matched only the red BSS sequence in M30. However, merging contact binaries (W UMa stars) were calculated to populate both sequences \citep{Kiraga2015}, consistent with the presence of known W UMa stars on both sequences \citep{Pietrukowicz2004}, as noted by \citetalias{Ferraro2009}. \citet{Jiang2017} showed that mass-transfer BSS products can populate the region of both sequences, which has been supported by a number of further calculations \citep{Lu10,Jiang22,Rain24,Wang2025}. None of these theoretical works predicted a clear gap matching the M30 observations. Observations of young white dwarf companions to BSSs in the open cluster NGC 188 show that distance of a BSS from the zero-age main sequence does not correlate closely with the age of the BSS \citep{Gosnell15}, arguing against simple interpretations of BSS features in the CMD.

The observational evidence for double BSS sequences remains limited to a handful of clusters, and 
uses different statistical tests in different clusters. 
Moreover, the interpretation of these features as evidence for two distinct formation mechanisms relies on the assumption that collisional and mass-transfer BSSs occupy systematically different regions of the CMD, an assumption that has not been thoroughly tested. 
In this context, we first conduct a systematic search for evidence of double BSS sequences in the largest homogeneous sample of high-quality cluster photometry available, the HUGS survey \citep{Sarajedini07, Piotto2015, Nardiello2018}. We then perform a detailed re-examination of the most prominent claimed detection of a double BSS sequence, NGC~7099 \citepalias{Ferraro2009}. 

In Section~\ref{sec:BSSacrosshugs}, we present our statistical analysis of BSS sequences across all 56 HUGS clusters using the Hartigan Dip Test \citep{Hartigan1985}, Akaike model comparisons between skewed unimodal and two-Gaussian models, and simulations accounting for realistic observed uncertainties. Section~\ref{sec:M30case} provides a detailed comparison of the double BSS sequence reported in NGC~7099 using three independent \textit{HST} datasets with different instruments and reduction methods. We discuss the implications of our findings in Section~\ref{sec:discussion} and summarize our conclusions in Section~\ref{sec:conclusion}.

\section{BSS\lowercase{s} across the HUGS clusters} \label{sec:BSSacrosshugs}

To perform a comprehensive analysis of the double BSS sequence across Milky Way GCs, we used the high-precision photometric and astrometric catalog produced by the \textit{Hubble Space Telescope} UV Globular Cluster Survey (HUGS; \citealt{Sarajedini07,Piotto2015, Nardiello2018}). This survey provides homogeneous ultraviolet and optical 
photometry 
for 56 Galactic globular clusters obtained with the WFC3/UVIS and ACS/WFC instruments. Optical data were collected in the \textit{F606W} and \textit{F814W} bands using ACS/WFC, while ultraviolet and blue observations in the \textit{F275W}, \textit{F336W}, and \textit{F438W} filters were obtained with WFC3/UVIS. A detailed summary 
is given by \citet{Sarajedini07,Piotto2015}. Data reduction followed the procedures described by \citet{Anderson2008} and \citet{Bellini2017}, and the final catalog provides three types of photometric measurements optimized for different stellar crowding regimes \citep{Nardiello2018}: Method~1 for bright, isolated stars; Method~2 for faint stars in moderately crowded fields; and Method~3 for highly crowded regions. Since our analysis focuses on the bright BSS population, we adopted Method~1 photometry. The astrometric calibration was tied to the Gaia~DR1 reference frame \citep{Gaia2016}, and cluster membership probabilities were estimated by comparing positions measured with ACS in one epoch, vs.\ WFC3 positions in an epoch 7-14 years later, using the local-sample method of \citet{Bellini2009}.

\subsection{HUGS Data filtering} \label{sec:hugs_filtering}

The Method~1 astro-photometric catalogs for all 56 GCs were retrieved from the HUGS archive\footnote{\url{https://archive.stsci.edu/prepds/hugs/}}. Cluster members were selected as stars with membership probabilities greater than 90\%. To ensure reliable photometry in the BSS selection (\S ~\ref{sec:bss_selection}), we considered only stars with \texttt{SHARP} values between $-0.15$ and $+0.15$ in both the \textit{F275W} and \textit{F336W} bands.

\citet{Legnardi2023} provides differential-reddening corrected HUGS photometry for the 21 GCs with the largest differential reddening effects. For these 21 GCs, we performed all further analysis on the differential-reddening corrected magnitudes. We note that none of the five HUGS GCs with the previously reported bifurcated BSSs have the differential-reddening corrected HUGS photometry available. 

\begin{figure*}
\centering
\includegraphics[width=1.8\columnwidth]{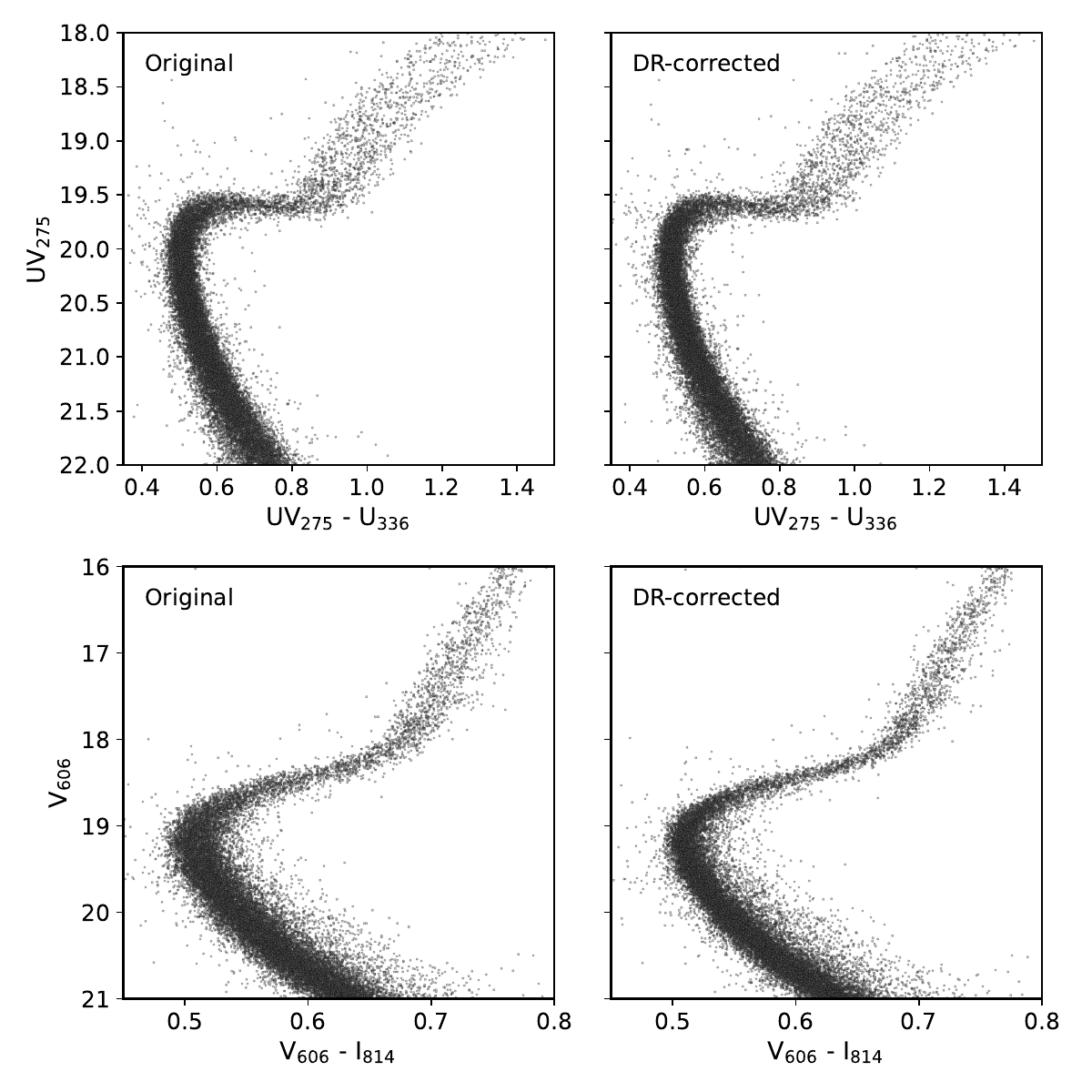}
\caption{HUGS UV$_{275}$--U$_{336}$ and V$_{606}$--I$_{814}$ CMDs of NGC~7078 before (left) and after (right) correcting for differential-reddening \citep{Legnardi2023}.}
\label{fig:ngc7078_diffred}
\end{figure*}

Among these five clusters, NGC~7078 (M15) has the highest reddening (by a factor of $\approx 2$). Therefore, for NGC~7078, we applied differential-reddening correction following the procedure of \citet{Legnardi2023} (see Figure~\ref{fig:ngc7078_diffred} and Section~\ref{sec:diptest_simulation}). All further analysis is performed on the differential-reddening corrected magnitudes for NGC~7078.

\subsection{BSS selection} \label{sec:bss_selection}

The selection of BSSs was carried out using the UV$_{275}$–U$_{336}$ (\textit{F275W}–\textit{F336W}) CMDs, where BSSs appear brighter, and are more easily distinguishable from the main sequence and from hot horizontal branch (HB) stars \citep{Raso2017}. The overall procedure follows the framework proposed by \citet{Raso2017}, with refinements to ensure uniformity across clusters with different metallicities, distances, and reddening values.

To achieve a homogeneous selection, we constructed a normalized CMD for each cluster, in which the main-sequence turnoff (MSTO) is placed at $m_{F275W}^* = 0$ and $(m_{F275W} - m_{F336W})^* = 0$. This normalization allows consistent comparison across clusters. Since the main-sequence turnoff (MSTO) and subgiant branch (SGB) morphologies vary with metallicity, we followed the approach adapted from \citet{Ferraro2018} to compute the shifts necessary for proper alignment:

\begin{enumerate}
\item The clusters were divided into five metallicity bins, spanning from [Fe/H] = -2.4 to -0.4, with each bin covering an interval of 0.4 dex ([Fe/H] values taken from \citet{Harris1996}).

\item For each metallicity group, we adopted a 12 Gyr isochrone with the corresponding composition from the BaSTI database\footnote{\url{http://basti-iac.oa-abruzzo.inaf.it/tracks.html}} \citep{Pietrinferni2021} and plotted it in the normalized CMD with the MSTO set at $(0, 0)$.

\item The isochrone corresponding to the metallicity group of each cluster was adjusted in reddening and distance to match the observed cluster principal sequences. 

\end{enumerate}

\begin{figure*}
    \centering
    \includegraphics[width=5.5cm]{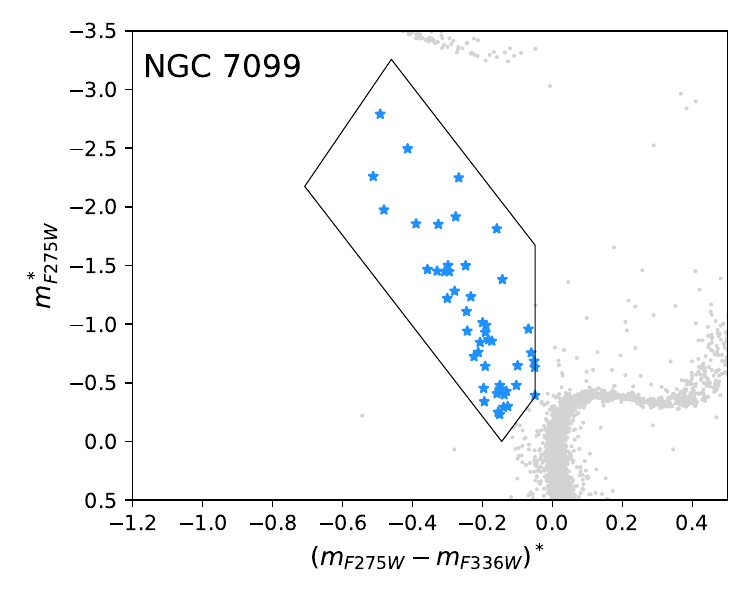}
    \includegraphics[width=5.5cm]{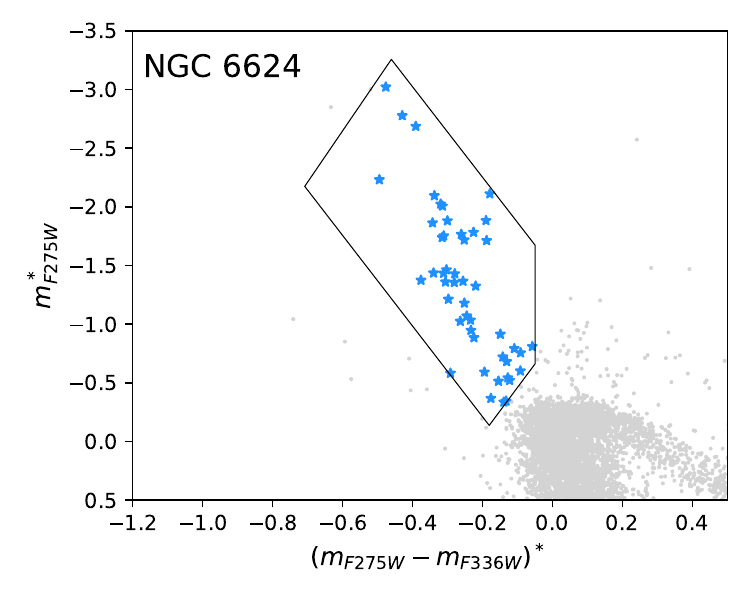}
    \includegraphics[width=5.5cm]{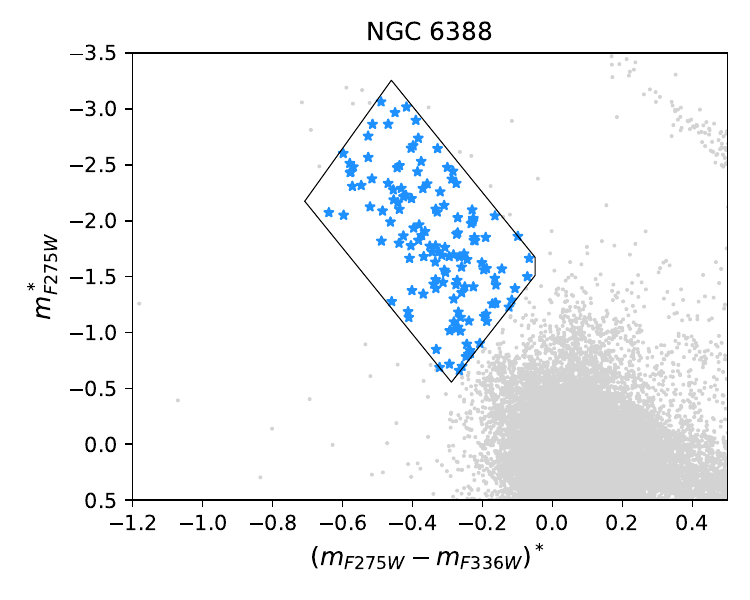}
    \caption{Normalized CMDs with the three different BSS selection regions used in this study. The selected BSSs are shown in blue, with the selection polygon indicated by solid black lines, and other cluster stars are shown in gray. The selection regions are adapted from \citet{Raso2017}. In NGC~7099, the red boundary separating fainter BSSs from the supra-MS plume follows the original definition by \citet{Raso2017}. In NGC~6624 and NGC~6388, we adopted two different red boundaries to avoid contamination from MS stars.
}
    \label{fig:bss_selections}
\end{figure*}

After normalization, BSSs were identified within the UV$_{275}$–U$_{336}$ CMDs using the selection boundaries defined by \citet{Raso2017}. For clusters where a tighter constraint was required to separate BSSs from stars belonging to the extended main-sequence plume, we refined the red boundary of the selection box to minimize contamination from MSTO stars (see Figure~\ref{fig:bss_selections}). As discussed by \citet{Raso2017}, the upper boundary of the selection box distinguishing very bright BSSs from HB stars is necessary; however, we found that introducing a blue boundary parallel to the red one was sufficient for this purpose in all GCs. We note that for all the GCs in which a double BSS sequence has been reported (see \S \ref{sec:intro}), no modification of the red boundary defined by \citet{Raso2017} was necessary.

\subsection{Dip Test and Akaike Model Comparison} \label{sec:hugs_diptest}

We employed the Hartigan Dip Test \citep{Hartigan1985} to assess the bimodality of the BSS sequences in the HUGS clusters (as done in \citetalias{Ferraro2009}). This test measures the maximum deviation between the empirical cumulative distribution and the best-fitting unimodal distribution, with the uniform distribution representing the null hypothesis. The dip statistic quantifies the deviation from unimodality, and the corresponding $p$-value indicates the significance of this deviation. A low $p$-value ($p \lesssim 0.05$) suggests evidence against unimodality, with smaller values corresponding to stronger evidence for bimodality, while a high p-value implies that the data are consistent with a unimodal distribution.

We applied the Hartigan Dip Test to the distributions of the shortest geometric distances of the BSSs in the UV$_{275}$--U$_{336}$ and V$_{606}$--I$_{814}$ CMDs from a 1~Gyr isochrone retrieved from the BaSTI database \citep{Pietrinferni2021}. We adopted the 1 Gyr isochrone as the reference because it provides a clear physical interpretation, representing a plausible evolutionary sequence for recently formed BSSs. We also repeated the Dip Test analysis using the 0.1~Gyr isochrone as the reference line for computing the BSS distance distribution, but this did not change our results significantly. The isochrone corresponding to the metallicity group of each cluster was adjusted to match the observed cluster photometry. 

For each BSS, the shortest signed geometric distance $d_{\mathrm{min}}$ to the isochrone was calculated as:

\begin{equation} \label{eq:geometric_distance}
d_{\mathrm{min}} = \pm\sqrt{(\mathrm{color}_{\mathrm{iso}} - \mathrm{color}_{\mathrm{BSS}})^2 + (\mathrm{mag}_{\mathrm{iso}} - \mathrm{mag}_{\mathrm{BSS}})^2}
\end{equation}

\noindent where $d_{\mathrm{min}}$ denotes the minimum distance evaluated over all points along the isochrone for a given BSS, with negative values assigned to BSSs lying to the left of the isochrone in the CMD. Color and magnitude refer to the CMD being analyzed.

\begin{figure*}
\centering
\includegraphics[width=1.8\columnwidth]{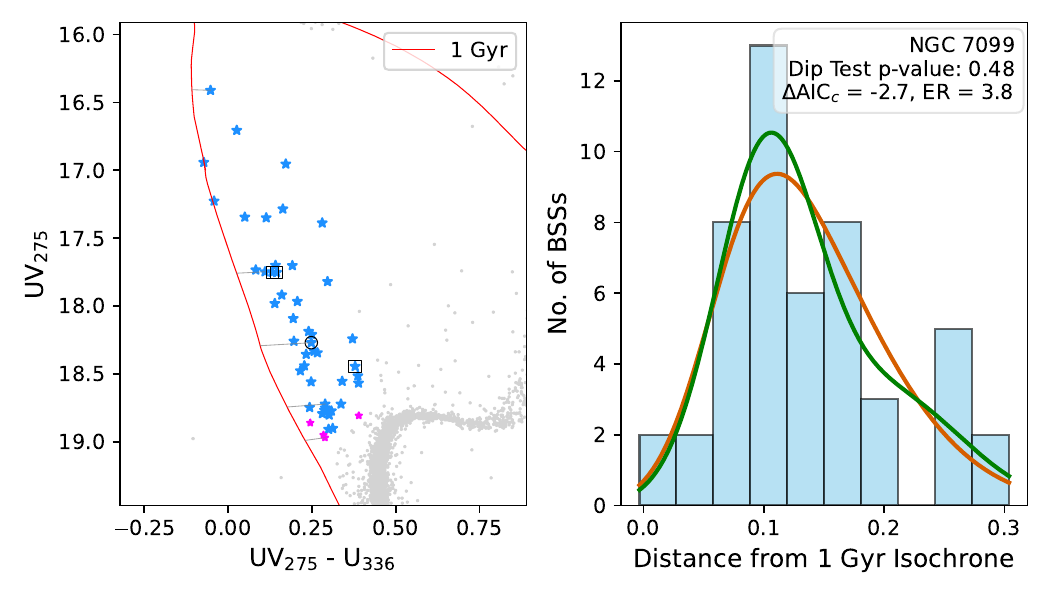}\\[1em]
\includegraphics[width=1.8\columnwidth]{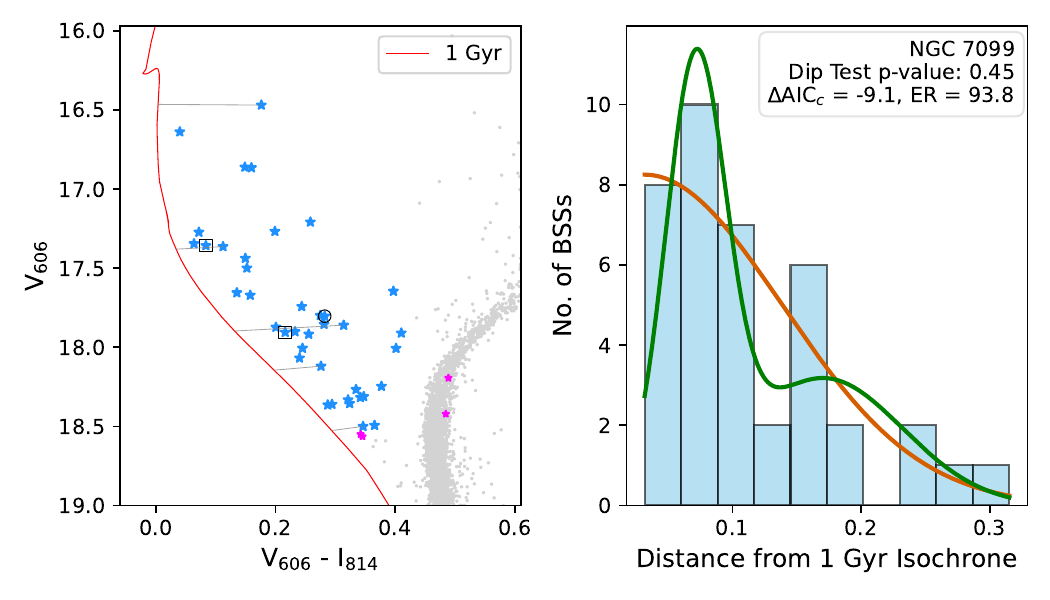}
\caption{Dip test analysis of the BSS sequence in the UV$_{275}$--U$_{336}$ and V$_{606}$--I$_{814}$ HUGS CMDs of NGC 7099 (M30). The red solid line shows the 1 Gyr BaSTI isochrone. Blue stars denote BSSs selected in the UV$_{275}$--U$_{336}$ that also occupy the BSS region in the optical CMD. Magenta stars show the BSSs (selected in UV$_{275}$--U$_{336}$) that do not occupy the BSS region in the optical CMD. Stars surrounded with squares are W~UMa-type stars, while those encircled with circles are SX~Phe variables, taken from \citet{Clement2001}. In the V$_{606}$--I$_{814}$ CMD, there is one less SX~Phe variable shown, as it didn't pass the optical photometry checks. The gray lines connecting some stars (at various magnitudes) to the 1~Gyr isochrone illustrate their shortest geometric distances in the CMD. The histograms show the distributions of the shortest geometric distances of the BSSs in the UV$_{275}$--U$_{336}$ and V$_{606}$--I$_{814}$ CMDs from the 1 Gyr isochrone. The fits of the skewed unimodal Gaussian model and the mixture of two unskewed Gaussian distributions are shown by the orange and green curves, respectively. The Hartigan dip test $p$-values, $\Delta\mathrm{AIC_c}$ and ER values of these distributions are given in the legends, for both UV$_{275}$--U$_{336}$ and V$_{606}$--I$_{814}$ BSS distributions.}
\label{fig:m30_hugs_diptest}
\end{figure*}

\begin{figure*}
    \centering

    \begin{subfigure}[t]{0.49\textwidth}
        \centering
        \includegraphics[width=\linewidth]{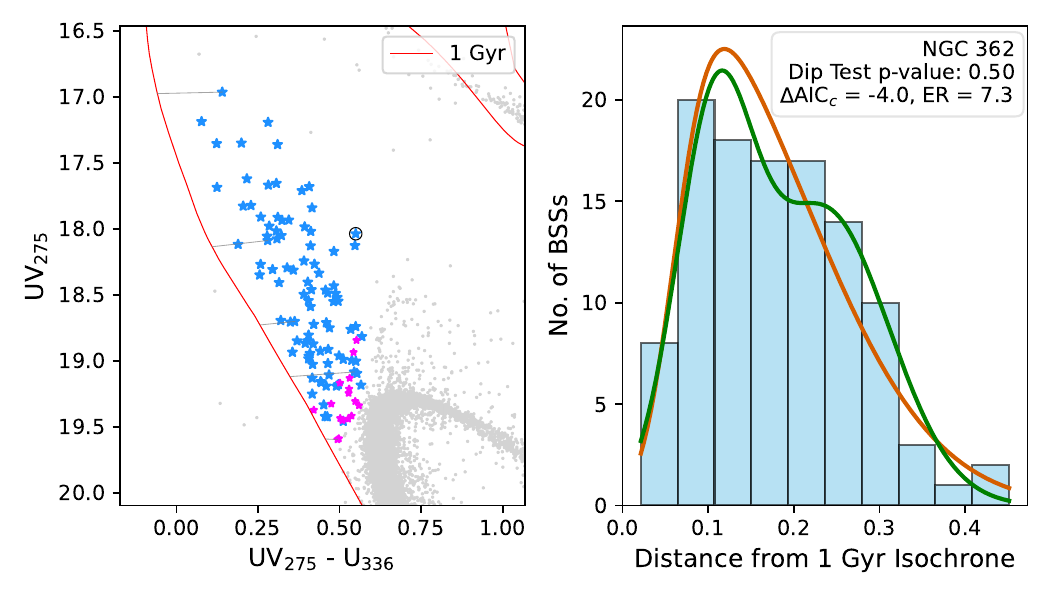}
    \end{subfigure}
    \hfill
    \begin{subfigure}[t]{0.49\textwidth}
        \centering
        \includegraphics[width=\linewidth]{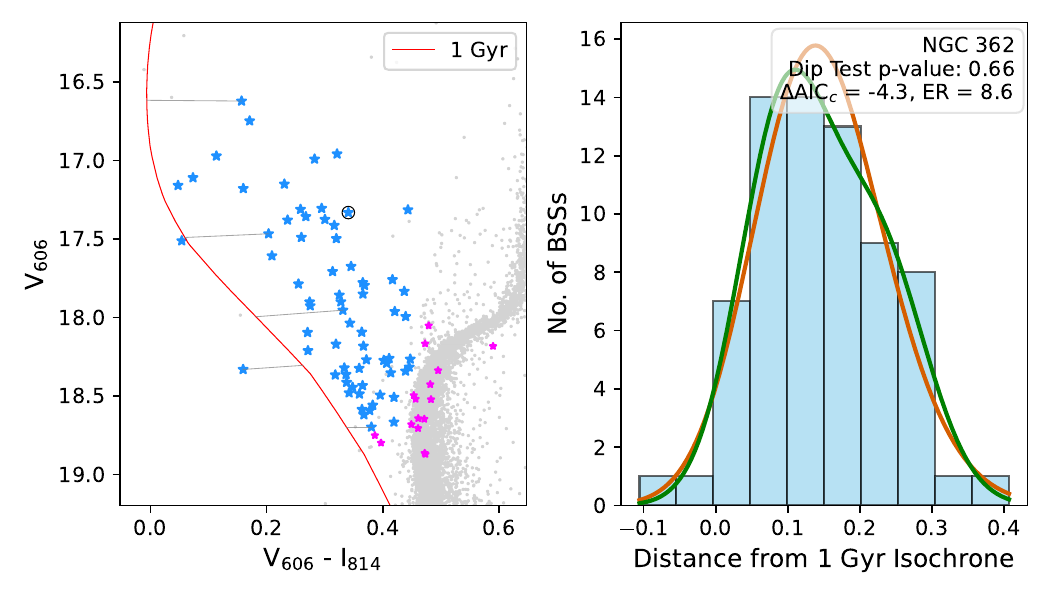}
    \end{subfigure}

    \begin{subfigure}[t]{0.49\textwidth}
        \centering
        \includegraphics[width=\linewidth]{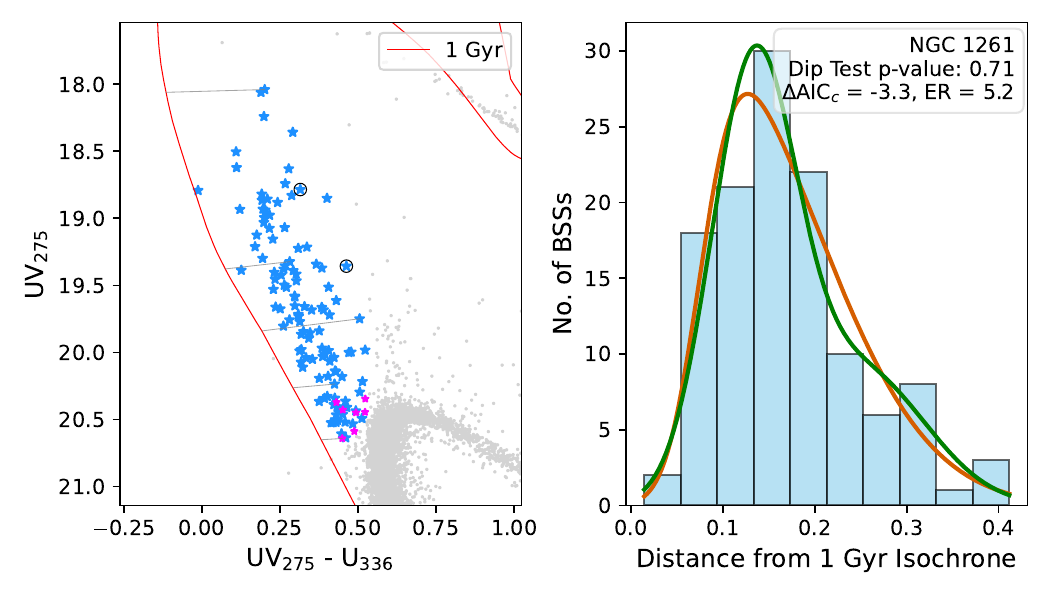}
    \end{subfigure}
    \hfill
    \begin{subfigure}[t]{0.49\textwidth}
        \centering
        \includegraphics[width=\linewidth]{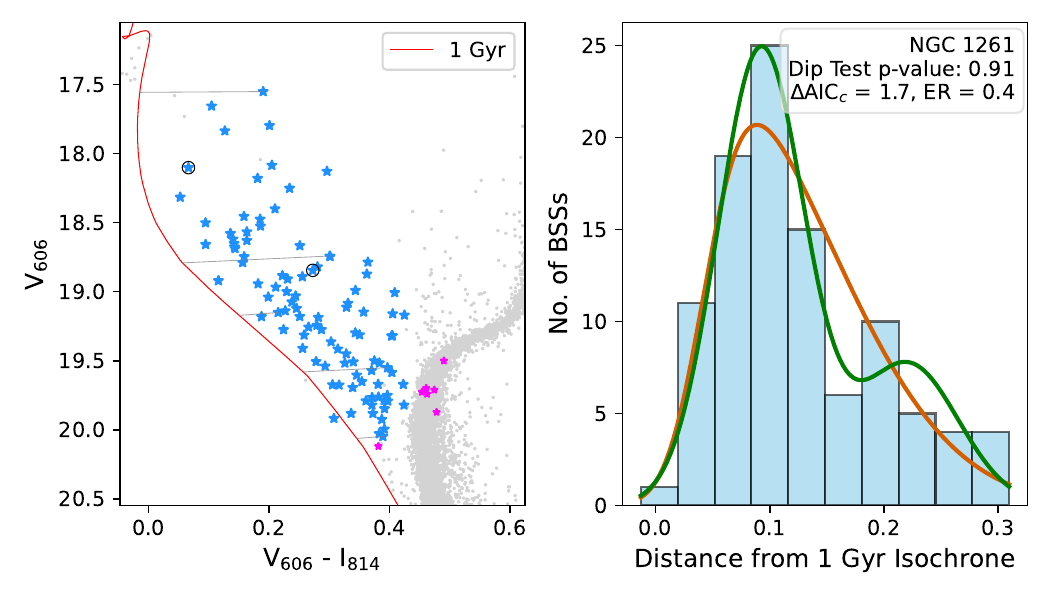}
    \end{subfigure}

    \begin{subfigure}[t]{0.49\textwidth}
        \centering
        \includegraphics[width=\linewidth]{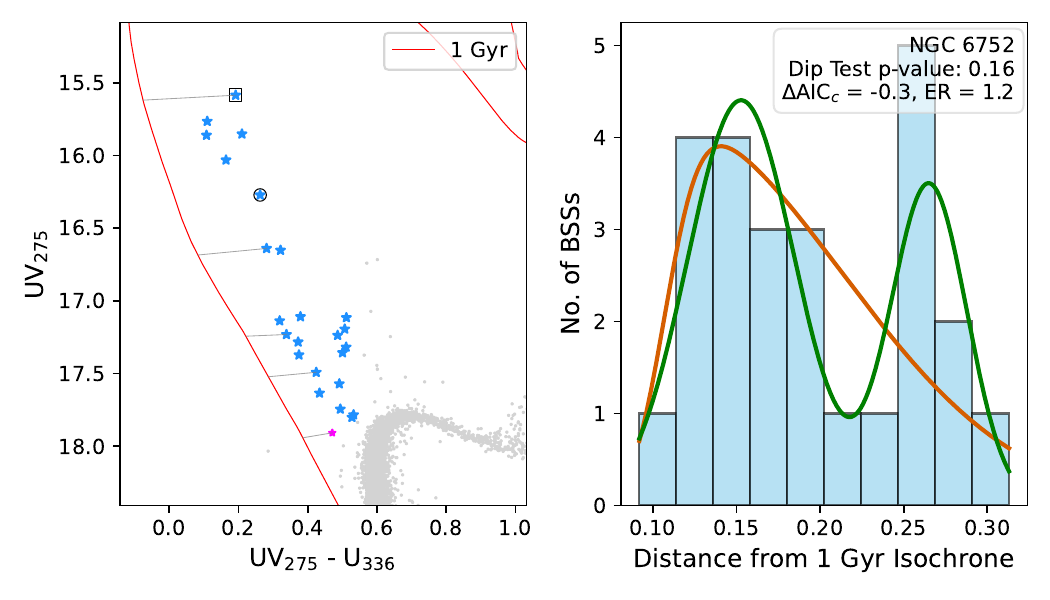}
    \end{subfigure}
    \hfill
    \begin{subfigure}[t]{0.49\textwidth}
        \centering
        \includegraphics[width=\linewidth]{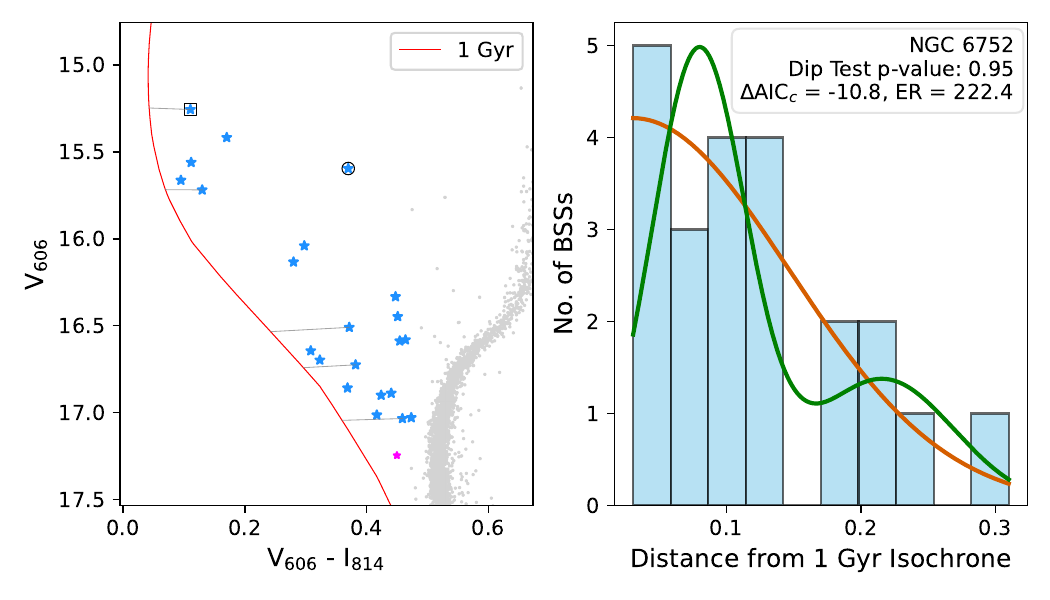}
    \end{subfigure}

    \begin{subfigure}[t]{0.49\textwidth}
        \centering
        \includegraphics[width=\linewidth]{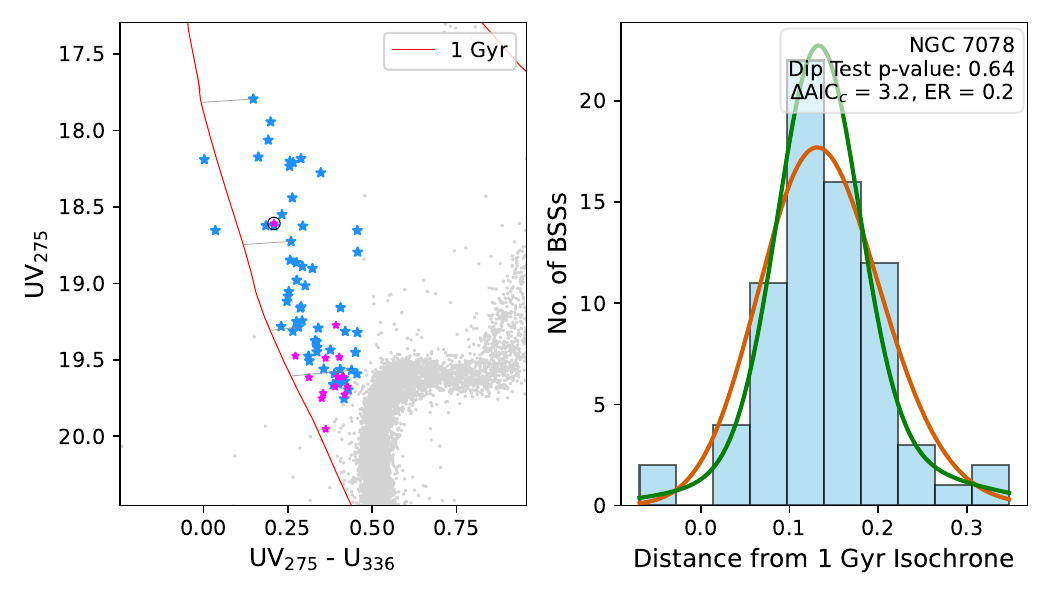}
    \end{subfigure}
    \hfill
    \begin{subfigure}[t]{0.49\textwidth}
        \centering
        \includegraphics[width=\linewidth]{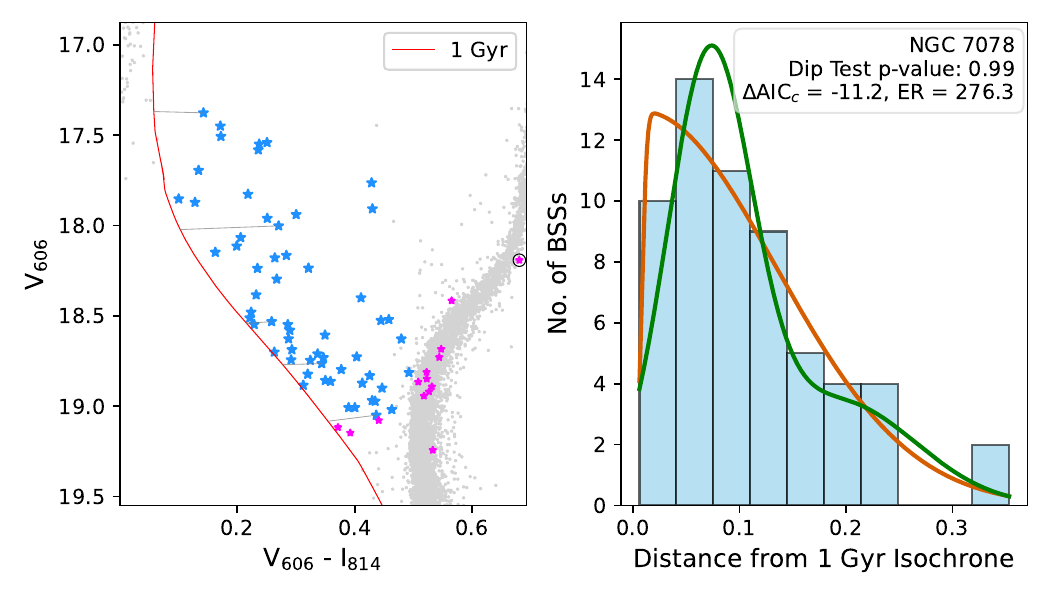}
    \end{subfigure}

    \caption{Dip test analysis of the BSS sequences in the UV$_{275}$--U$_{336}$ and V$_{606}$--I$_{814}$ CMDs of HUGS clusters in which a double BSS sequence has been previously reported (NGC~362, NGC~1261, NGC~6752, and NGC~7078). The red solid line shows the 1 Gyr BaSTI isochrone. Blue stars denote BSSs selected in the UV$_{275}$--U$_{336}$ that also occupy the BSS region in the optical CMD. Magenta stars show the BSSs (selected in UV$_{275}$--U$_{336}$) that do not occupy the BSS region in the optical CMD. 
    Stars surrounded with squares are W~UMa-type stars, while those encircled with circles are SX~Phe variables, taken from \citet{Clement2001}. The gray lines connecting some stars (at various magnitudes) to the 1~Gyr isochrone illustrate their shortest geometric distances in the CMD. The histograms show the distributions of the shortest geometric distances of the BSSs in the UV$_{275}$--U$_{336}$ and V$_{606}$--I$_{814}$ CMDs from the 1 Gyr isochrone. The fits of the skewed unimodal Gaussian model and the mixture of two unskewed Gaussian distributions are shown by the orange and green curves, respectively. The Hartigan dip test $p$-values, $\Delta\mathrm{AIC_c}$ and ER values of these distributions are given in the legends, for both UV$_{275}$--U$_{336}$ and V$_{606}$--I$_{814}$ BSS distributions.}
    \label{fig:diptest_hugs_4gc}
\end{figure*}

The BSSs were selected from the UV$_{275}$--U$_{336}$ CMD (see \S~\ref{sec:bss_selection}). These selected stars were then plotted in the V$_{606}$--I$_{814}$ CMD after applying the photometric quality criterion that \texttt{SHARP} values in both the $F606W$ and $F814W$ bands lie between $-0.15$ and $+0.15$. We found that some of the stars identified as BSSs in the UV$_{275}$--U$_{336}$ CMD do not occupy the same region in the V$_{606}$--I$_{814}$ CMD (see Figure~\ref{fig:m30_hugs_diptest} and \ref{fig:diptest_hugs_4gc}). Therefore, we retained only those stars in the V$_{606}$--I$_{814}$ CMD that are both brighter and bluer than the MSTO as BSSs.

\begin{figure*}
\centering
\includegraphics[width=1.8\columnwidth]{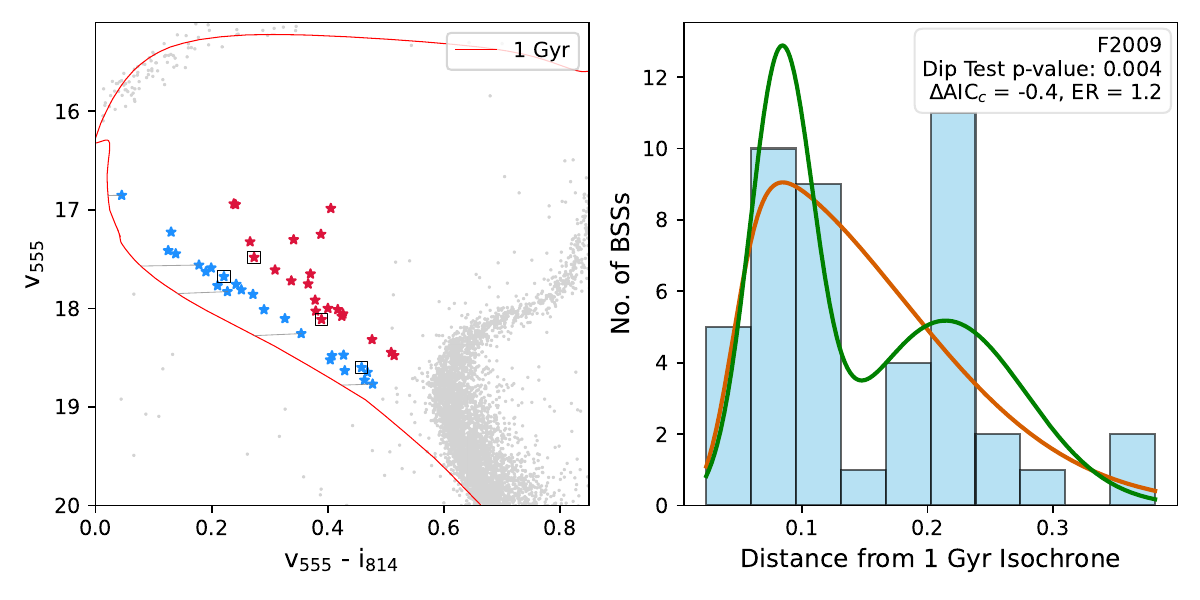}\\[1em]
\includegraphics[width=1.8\columnwidth]{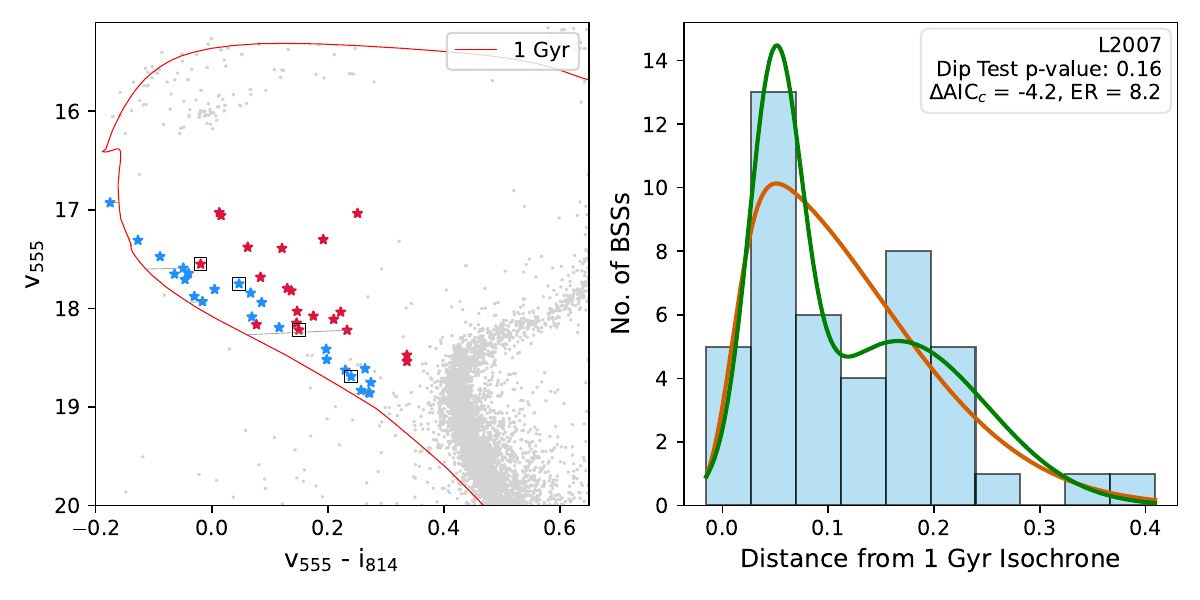}
\caption{Dip test analysis of the BSS sequence in the v$_{555}$--i$_{814}$ HST/WFPC2 CMD of NGC~7099 (M30) from \citetalias{Ferraro2009} (top) and \citetalias{Lugger2007} (bottom). The red solid line shows the 1~Gyr BaSTI isochrone. The BSSs are color-coded as blue and red based on the classification by \citetalias{Ferraro2009}. The \citetalias{Lugger2007} CMD shows the same cross-matched BSSs, except for one BSS not found in the \citetalias{Lugger2007} catalog. Stars surrounded with squares are W~UMa-type stars, taken from \citet{Clement2001}.  The gray lines connecting some stars (at various magnitudes) to the 1~Gyr isochrone illustrate their shortest geometric distances in the CMD. The histograms show the distributions of the shortest geometric distances of the BSSs from the 1~Gyr isochrone. The fits of the skewed unimodal Gaussian model and the mixture of two unskewed Gaussian distributions are shown by the orange and green curves, respectively. The Hartigan dip test $p$-values, $\Delta\mathrm{AIC_c}$ and ER values of these distributions are given in the legends.}
\label{fig:m30_wfpc2_diptest}
\end{figure*}

We also searched for variable stars that may be contaminating the BSS sequence in the HUGS clusters for which a double BSS sequence has been previously reported (see Figures~\ref{fig:m30_hugs_diptest}, \ref{fig:diptest_hugs_4gc}, and \ref{fig:m30_wfpc2_diptest}). We cross-matched the selected BSSs in these clusters with the Catalogue of Variable Stars in Galactic Globular Clusters\footnote{\url{https://www.astro.utoronto.ca/~cclement/read.html}} \citep{Clement2001}. The two categories where variability can affect the BSS sequence are SX~Phe variables, which exhibit short-period pulsations on timescales of $\lesssim 0.2$~days, and W~UMa-type stars, contact binaries with periods concentrated around 0.25--0.6~days \citep{Rucinski00}.

No BSS distribution exhibited a Dip Test $p$-value lower than 0.15.  Table~\ref{tab:diptest_hugs} lists the $p$-values for 5 of the 7 clusters  in which a double BSS sequence has been previously reported. (NGC~6256 and NGC~2173 are not included in the HUGS dataset.) 

\citet{Dalessandro2013} compared the distribution of distances of BSSs in NGC 362 using Gaussian mixture modeling \citep{Muratov2010}, finding that a bimodal Gaussian model provided a significantly better fit than a single Gaussian. However, this assumes that the distribution is symmetric in color. The histograms of the BSS distance distributions appear to be better modeled by a skewed unimodal Gaussian (\texttt{scipy.stats.skewnorm}\footnote{\url{https://docs.scipy.org/doc/scipy/reference/generated/scipy.stats.skewnorm.html}}) than by a mixture of two unskewed Gaussian distributions (\texttt{sklearn.mixture.GaussianMixture}\footnote{\url{https://scikit-learn.org/stable/modules/generated/sklearn.mixture.GaussianMixture.html}}) (see \S~\ref{sec:discussion}). To test this quantitatively, we compared the Akaike Information Criterion with finite sample correction ($\mathrm{AIC_c}$, \citet{AIC_2002, Wagenmakers2004}) for a skewed unimodal Gaussian model ($\mathrm{AIC_{c,skew}}$) vs. a mixture of two unskewed Gaussians ($\mathrm{AIC_{c,gmm}}$), fitted to the BSS distance distributions. Because the number of BSSs in each cluster is relatively small ($N < 200$), we adopted $\mathrm{AIC_c}$ rather than the standard Akaike or Bayesian Information Criteria (AIC/BIC). The $\mathrm{AIC_c}$ is recommended for smaller data sets, particularly when $N/k < 40$ \citep{AIC_2002}.

The $\mathrm{AIC_c}$ is defined in Equation~\ref{eq:aic}, where $\mathcal{L}_{\mathrm{max}}$ is the maximum likelihood achievable by the model, $k$ is the number of free parameters, and $N$ is the number of data points \citep{AIC_2002, Wagenmakers2004}: 

\begin{equation} \label{eq:aic}
\mathrm{AIC_c} = -2 \ln \mathcal{L}_{\mathrm{max}} + 2 k + \frac{2 k (k+1)}{N-k-1}
\end{equation}

The model with lower $\mathrm{AIC_c}$ value is considered the better fit. We calculated $\Delta\mathrm{AIC_c} = \mathrm{AIC_{c,skew}} - \mathrm{AIC_{c,gmm}}$ and the corresponding Evidence Ratio (ER) of Akaike Weights for the distance distributions in both the UV$_{275}$--U$_{336}$ and V$_{606}$--I$_{814}$ CMDs across all 56 HUGS clusters. The Evidence Ratio provides a more intuitive interpretation: a ER of $x$ indicates that the skewed unimodal Gaussian model is $x$ times more likely than the mixture of two unskewed Gaussian distributions, given the data. The ER can be calculated from $\Delta\mathrm{AIC_c}$ using Equation~\ref{eq:er} \citep{Wagenmakers2004}: 

\begin{equation} \label{eq:er}
    \mathrm{ER} = \exp\left(-\frac{1}{2}\Delta\mathrm{AIC_c}\right)
\end{equation}

ER $> 1$ indicates support for the skewed unimodal Gaussian model over the mixture of two unskewed Gaussian distributions, with larger values implying stronger support for the skewed unimodal model. Conversely, $\mathrm{ER} < 1$ indicates support for the mixture of two unskewed Gaussians, with smaller values implying stronger evidence for this model over the skewed unimodal Gaussian model.

Across all 56 HUGS clusters, 7 and 11 clusters in the UV$_{275}$--U$_{336}$ and V$_{606}$--I$_{814}$ CMDs, respectively, show $\mathrm{ER} < 1$, indicating that a mixture of two unskewed Gaussian distributions provides a better fit than the skewed unimodal model.  Among these 18 cases, 16 have $\mathrm{ER} \geq 0.18$, with the other two at 0.03 and 0.05; so the evidence for two Gaussians is not strong. Finding only 18 cases with $\mathrm{ER} < 1$ out of 112 total indicates the overall preference for the skewed unimodal model. Furthermore, in nearly all of these 18 cases, the mixture of two unskewed Gaussians performs better solely because of a secondary peak at larger distances containing only $\sim$5--10 BSSs, which is inconsistent with the reported double sequence morphology of a narrower blue and a broader red sequence containing roughly equal numbers of stars. 

Notably, among the HUGS clusters with previously reported double BSS sequences, only NGC~1261 (V$_{606}$--I$_{814}$ CMD) and NGC~7078 (UV$_{275}$--U$_{336}$ CMD) show ER $< 1$, favoring a bimodal model. For NGC~1261, the apparent secondary peak in the V$_{606}$--I$_{814}$ CMD comprises only $\sim$20 BSSs out of $\sim$100 total (see Figure~\ref{fig:diptest_hugs_4gc}); this feature does not match the suggested morphology of a bifurcated BSS sequence. For NGC~7078 in the UV$_{275}$--U$_{336}$ CMD (Figure~\ref{fig:diptest_hugs_4gc}), the distribution shows no compelling signature of bimodality. We suspect that the preference of AIC$_c$ for bimodality in this case reflects the fit creating two closely-spaced peaks, to produce a broader central peak with sharper edges than a single Gaussian (note that the green line in Fig. 4 falls off faster on both sides). This shape, again, does not match the suggested morphology of a bifurcated BSS sequence, as it lacks any evidence for a gap between sequences.
All remaining HUGS clusters in which a double BSS sequence has been previously reported exhibit $\mathrm{ER}$ values favoring the skewed unimodal Gaussian model over the mixture of two unskewed Gaussian distributions (see Table~\ref{tab:diptest_hugs}, Figures~\ref{fig:m30_hugs_diptest} \& \ref{fig:diptest_hugs_4gc}).

\subsection{Observational Error and Dip Test simulation} \label{sec:diptest_simulation}

We investigated the influence of observational uncertainties on the detectability of a bifurcated BSS sequence and assessed whether the Hartigan Dip Test \citep{Hartigan1985} can reliably identify bimodality once the intrinsic distribution is smeared by observed photometric errors. The sub-giant branch (SGB) width was adopted as a proxy for the observational error, since SGB stars generally have magnitudes comparable to those of BSSs, and the SGB is the sharpest sequence in these CMDs. While some clusters (e.g. $\omega$ Cen, M22, NGC 7089) show multiple subgiant branches in UV$_{275}$--U$_{336}$, these clusters are few, and do not have prior claims of  bifurcated BSS sequences.

\begin{figure}
  \resizebox{\hsize}{!}{\includegraphics{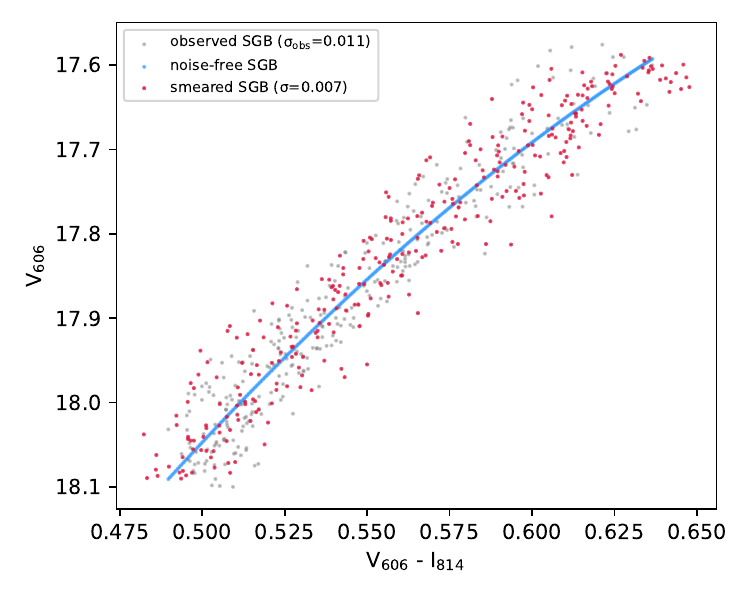}}
  \caption{Example showing the estimation of observed uncertainty ($\sigma$) from the observed SGB width in the V$_{606}$–I$_{814}$ CMD of NGC~7099.  The gray dots indicate observed SGB stars, the blue line represents the input simulated SGB stars lying on the fitted spline to the SGB, and the red dots represent simulated SGB stars with $\sigma=0.007$ applied. }
  \label{fig:sgb_width}
\end{figure}

We estimated the effective observational uncertainty ($\sigma$) from the observed SGB width as follows: a smooth spline was fitted to the SGB in the CMDs, representing an idealized, noise-free SGB. The perpendicular distances of individual SGB stars from this spline were computed to quantify the observed SGB width, and the resulting distribution of perpendicular distances was fitted with a Gaussian to determine the observed dispersion ($\sigma_{\mathrm{obs}}$). We then created a synthetic SGB based on the spline fit and iteratively convolved it with Gaussian noise of varying width, assuming the same trial Gaussian noise values in both filters ($\sigma_V = \sigma_I$ for the V$_{606}$--I$_{814}$ CMD, and similarly for the UV$_{275}$--U$_{336}$ CMD). The Gaussian width was adjusted until the dispersion of the smeared synthetic SGB matched $\sigma_{\mathrm{obs}}$. The applied Gaussian noise value was adopted as the effective observed uncertainty ($\sigma$) per filter in the SGB region, which we then applied to the BSS sequence in our simulations. The observed uncertainty ($\sigma$) includes contributions from photometric uncertainty, differential reddening, multiple stellar populations, and other observational effects. Figure~\ref{fig:sgb_width} illustrates an example of this $\sigma$ estimation procedure.

To evaluate the dip test’s sensitivity to bimodality under realistic conditions, we constructed a model BSS distribution consisting of two infinitely sharp sequences, with 21 and 24 stars representing the red and blue BSSs reported by \citetalias{Ferraro2009} in M30, located at normalized positions 0.01 and 0.11, respectively. From this model, we generated 1000 random samples per cluster, each with the same number of BSSs as observed in each cluster (or a maximum of 45, matching the input model, for richer clusters). Each sample was then perturbed by adding Gaussian noise with a standard deviation equal to the estimated observational uncertainty ($\sigma$) derived from the corresponding SGB.

We applied Hartigan’s Dip Test \citep{Hartigan1985} to each simulated sample to assess the detectability of bimodality, adopting $p < 0.05$ as the criterion for significant evidence of a double sequence. We found that all 56 clusters in the V$_{606}$–I$_{814}$ CMDs exhibited detectability rates above 50\%. However, in the UV$_{275}$–U$_{336}$ CMDs, only 10 out of 56 clusters show detectability rates above 50\%. This suggests that, even after accounting for realistic observed uncertainties and limited BSS sample sizes, the dip test should be able to identify a genuine double BSS sequence in a significant fraction of HUGS clusters if such bimodality truly exists, at least using V$_{606}$–I$_{814}$ CMDs. The lower detectability rates in the UV$_{275}$–U$_{336}$ CMDs are likely due to higher observational uncertainties ($\sigma$) derived from the SGB width, which in turn reflects the lower photon counts per source typically obtained in the $F275W$ and $F336W$ filters. Table~\ref{tab:diptest_hugs} lists the observational uncertainties ($\sigma$) and the detectability rates for the HUGS GCs in which a double BSS sequence has been previously reported.

The effect of differential reddening in NGC~7078 (see Figure~\ref{fig:ngc7078_diffred}) is also evident from the observational uncertainties ($\sigma$). We found that the SGB width (a proxy for observational uncertainty, $\sigma$) decreased from $0.010$ and $0.061$ to $0.006$ and $0.051$ in the V$_{606}$--I$_{814}$ and UV$_{275}$--U$_{336}$ CMDs, respectively, after applying a differential-reddening correction. The smaller sharpening of the UV$_{275}$--U$_{336}$ CMD with this correction may be explained by the effects of multiple populations causing broadening in the UV$_{275}$--U$_{336}$ color \citep{Piotto2015}.

\begin{deluxetable}{lcccccccccc}
\tablecaption{Double BSS sequence analysis for HUGS clusters in which a double BSS sequence has been previously reported. Hartigan dip test $p$-values, $\Delta\mathrm{AIC_c} = \mathrm{AIC_{c,skew}} - \mathrm{AIC_{c,gmm}}$, and Evidence Ratio (ER) are provided for both UV$_{275}$--U$_{336}$ and V$_{606}$--I$_{814}$ CMDs (see \S \ref{sec:hugs_diptest}). The SGB width, as a proxy of observational error, is given as $\sigma$. The ``Detect. Rate'' provides the detectability rate for a double BSS sequence (see \S \ref{sec:diptest_simulation}). \label{tab:diptest_hugs}}
\tablehead{
\colhead{Cluster} &
\multicolumn{5}{c}{UV$_{275}$–U$_{336}$} &
\multicolumn{5}{c}{V$_{606}$–I$_{814}$} \\
\colhead{} &
\colhead{$p$-value} &
$\Delta\mathrm{AIC_c}$ &
ER & 
\colhead{$\sigma$} &
\colhead{Detect. Rate} &
\colhead{$p$-value} &
$\Delta\mathrm{AIC_c}$ &
ER &
\colhead{$\sigma$} &
\colhead{Detect. Rate}}

\startdata
NGC~362  & 0.50 & -4.0 & 7.3  & 0.013 & 100 \% & 0.66 & -4.3 & 8.6  &  0.005 & 100 \%  \\
NGC~1261  & 0.71 & -3.3 & 5.2  & 0.021  & 72 \% & 0.91 & 1.7 & 0.4  &  0.005 & 100 \%  \\
NGC~6752 & 0.16 & -0.3 & 1.2  & 0.024 & 25 \% & 0.95 & -10.8 & 222  &  0.006 & 98 \%  \\
NGC~7078 & 0.64 & 3.2 & 0.2  & 0.051 & 1 \% & 0.99 & -11.2 & 276  &  0.006 & 100 \%  \\
NGC~7099 & 0.48 & -2.7 & 3.8  & 0.020 & 80 \% & 0.45 & -9.1 & 94  &  0.007 & 100 \%  \\
\enddata
\end{deluxetable}

\section{The case of NGC~7099 (M30)} \label{sec:M30case}

NGC~7099 (M30) represents the first and most compelling case in which a double BSS sequence has been reported in the literature \citepalias{Ferraro2009}. In this section, we revisit the reported bifurcation using the original \textit{HST} WFPC2 photometry presented by \citetalias{Ferraro2009}, as well as an independent WFPC2 reduction from \citet{Lugger2007} (hereafter \citetalias{Lugger2007}). We further examine the BSS sequence using the newer \textit{HST} WFC3/UVIS and ACS/WFC HUGS datasets. Together, these datasets allow us to directly test whether the double BSS sequence persists across instruments, filters, and photometric reduction methods.

\subsection{WFPC2 Observations}

The core of NGC~7099 (M30) was extensively observed with \textit{HST}/WFPC2 in 1999 as part of Program GO-7379 (PI: Edmonds), using the u$_{336}$ ($F336W$), v$_{555}$ ($F555W$), and i$_{814}$ ($F814W$) filters. These high-quality WFPC2 observations form the foundation of several photometric studies of the stellar populations in M30 \citep{Pietrukowicz2004, Lugger2007, Ferraro2009}.

\subsection{Ferraro+09} \label{sec:f09}

We re-analyzed the double BSS sequence reported by \citetalias{Ferraro2009} using their photometry. Figure~\ref{fig:m30_wfpc2_diptest} shows the dip test analysis applied to the BSSs identified in the v$_{555}$--i$_{814}$ CMD by them. Following the procedure described in \S~\ref{sec:hugs_diptest}, we applied the Hartigan Dip Test \citep{Hartigan1985} to the distribution of shortest geometric distances of the BSSs from a 1~Gyr BaSTI isochrone \citep{Pietrinferni2021} corresponding to the metallicity of NGC~7099 ([Fe/H] = $-2.27$; \citealt{Harris1996}), adjusted in distance and reddening to match the observed cluster photometry. 
We obtained a $p$-value of $4\times10^{-3}$, indicating moderate ($\sim$3$\sigma$) evidence for bimodality. However, \citetalias{Ferraro2009} reported a substantially lower $p$-value of $\approx 1 \times 10^{-5}$, though they used a straight line fitted to the blue BSS sequence as the reference for computing distances, rather than an isochrone. We attempted to reproduce the dip test analysis using \citetalias{Ferraro2009}'s method of calculating perpendicular distances from a straight line fitted to the blue BSS sequence. Note that for a straight line, the shortest geometric distance (as defined in Equation~\ref{eq:geometric_distance}) is mathematically equivalent to the perpendicular distance, if the reference is a straight line, so both approaches should yield identical results. Despite this, we obtained a $p$-value of $2 \times 10^{-3}$, significantly higher than their reported value of $\approx 1 \times 10^{-5}$. F.~R. Ferraro found (priv. comm.) a p-value of $1.5\times10^{-3}$ when re-analyzing the \citetalias{Ferraro2009} datapoints with the dip test. 

While these updated p-values (between $1.5\times10^{-3}$ and $4\times10^{-3}$, depending on the exact test) may be significant for a single test, they become less so when the number of possible trials is included. The correct number of trials to compare with is not obvious, but might include the number of CMDs investigated and/or the range of potential unusual CMD features that might be seen (often termed the look-elsewhere effect; see e.g. \citealt{BayerSeljak20}). For example, if one cluster of the 56 in HUGS shows a feature with a p-value of $1.5\times10^{-3}$, the probability of finding one when searching all HUGS clusters is $1-(1-1.5\times10^{-3})^{56}=8.1$\%.

We also compared the $\Delta\mathrm{AIC_c}$ and ER for fits to the 
distribution of shortest geometric distances of the BSSs from the 1~Gyr 
isochrone (see Figure~\ref{fig:m30_wfpc2_diptest}) using a skewed unimodal 
Gaussian model ($\mathrm{AIC_{c,skew}}$), vs.\ a mixture of two unskewed 
Gaussian distributions ($\mathrm{AIC_{c,gmm}}$). We obtained 
$\Delta\mathrm{AIC_c} = -0.4$ and $\mathrm{ER} \approx 1.2$ 
(see \S~\ref{sec:hugs_diptest}), indicating only marginal preference for the skewed unimodal Gaussian model. Using \citetalias{Ferraro2009}'s method of calculating distances from a straight line fitted to the blue BSS sequence, we obtained $\Delta\mathrm{AIC_c} = -0.5$ and $\mathrm{ER} \approx 1.6$, again showing only a marginal preference for the skewed unimodal Gaussian model over the mixture of two unskewed Gaussians.

Using the method described in \S~\ref{sec:diptest_simulation}, we estimated the observational uncertainty to be $\sigma = 0.020$ for the \citetalias{Ferraro2009} v$_{555}$--i$_{814}$ CMD. This value reflects the intrinsic observational scatter derived from the subgiant branch width and does not indicate any problem with the quality of the \citetalias{Ferraro2009} photometry itself. Our simulations indicate that a dip test $p$-value below 0.05 should be detected in 79\% of 1000 simulated samples, suggesting that bimodality should be generally detectable if present. However, the ability to detect an extremely sharply peaked bimodal distribution, generating $p < 1 \times 10^{-5}$, as originally reported by \citetalias{Ferraro2009}, drops to only 3\% (or 31\% for a $p < 1.5 \times 10^{-3}$).

\subsection{Lugger+07} \label{sec:L07}

\citetalias{Lugger2007} carried out a photometric analysis of the \emph{HST} GO-7379 data set using aperture photometry on drizzle-combined frames. We briefly summarize their approach, which was performed using IRAF/STSDAS routines. The observations consist of four exposures in each of the F555W and F814W filters and three exposures in F336W at each of 12 dither positions, yielding a total of 46 F555W, 48 F814W, and 38 F336W images. For each filter and dither position, the individual exposures were first stacked using {\tt combine}. To cover the area beyond the PC chip, the four WFPC2 detectors were merged into a single mosaic with the {\tt wmosaic} routine, producing one mosaic per dither point. The resulting 12 stacked PC frames and 12 WFPC2 mosaics were then drizzle-combined with $2\times$ oversampling to enhance the effective spatial resolution. Aperture photometry was performed on both the drizzle-reconstructed PC frame and the outer-region mosaic using a 4-pixel aperture (in oversampled units). The final photometric catalog adopted measurements from the PC chip for the cluster core and from the surrounding WFPC2 mosaic for the outer regions.  We note that this photometry is not designed to accurately capture the brightest stars ($V<16.5$). 

We cross-matched the 45 BSSs reported by \citetalias{Ferraro2009} in M30 with the independent WFPC2 photometry of \citetalias{Lugger2007}, successfully identifying 44 matches. The one unmatched BSS (B16) was found to be an extended object with zero HUGS membership probability, suggesting it is not a genuine cluster member (see Figure \ref{fig:findingcharts}). Following the procedure described in \S~\ref{sec:f09}, we applied the dip test analysis to these 44 BSSs in the v$_{555}$--i$_{814}$ CMD of \citetalias{Lugger2007}. We obtained a $p$-value of 0.162, indicating no significant evidence for bimodality, consistent with visual inspection of the CMD (see Figure~\ref{fig:m30_wfpc2_diptest}). Following the procedure described in \S~\ref{sec:hugs_diptest} and \ref{sec:f09}, we found $\Delta\mathrm{AIC_c} = -4.2$ and $\mathrm{ER} \approx 8$ (see Figure~\ref{fig:m30_wfpc2_diptest}), indicating positive evidence favoring the skewed unimodal Gaussian model over the mixture of two unskewed Gaussian distributions.

Using the method described in \S~\ref{sec:diptest_simulation}, we estimated the observational uncertainty to be $\sigma = 0.023$ for the \citetalias{Lugger2007} v$_{555}$--i$_{814}$ CMD, comparable to (but slightly larger than) that of \citetalias{Ferraro2009} ($\sigma = 0.020$). Our simulations indicate that a dip test $p$-value below 0.05 should be detected in 55\% of realizations, suggesting a decent chance that bimodality would be detectable.

We also calculated the observational uncertainty from the width of the main sequence (MS) for both the \citetalias{Ferraro2009} and \citetalias{Lugger2007} v$_{555}$--i$_{814}$ CMDs using a method similar to that described in \S~\ref{sec:diptest_simulation}, applied to the MS (using a box of about $\sim$0.1 in color and 2.5 in magnitude). The observational uncertainties derived from the MS width were found to be $\sigma = 0.026$ and $\sigma = 0.029$ for \citetalias{Ferraro2009} and \citetalias{Lugger2007}, respectively. These values are larger than those derived from the SGB width. However, they demonstrate that both photometric datasets are comparable even when using the MS width as the uncertainty estimate. We note that the MS is intrinsically broader than the SGB (both due to unresolved binaries and due to its fainter magnitudes), so the SGB-derived uncertainties are likely more accurate for estimating BSS observed uncertainties.

\begin{figure}
  \resizebox{\hsize}{!}{\includegraphics{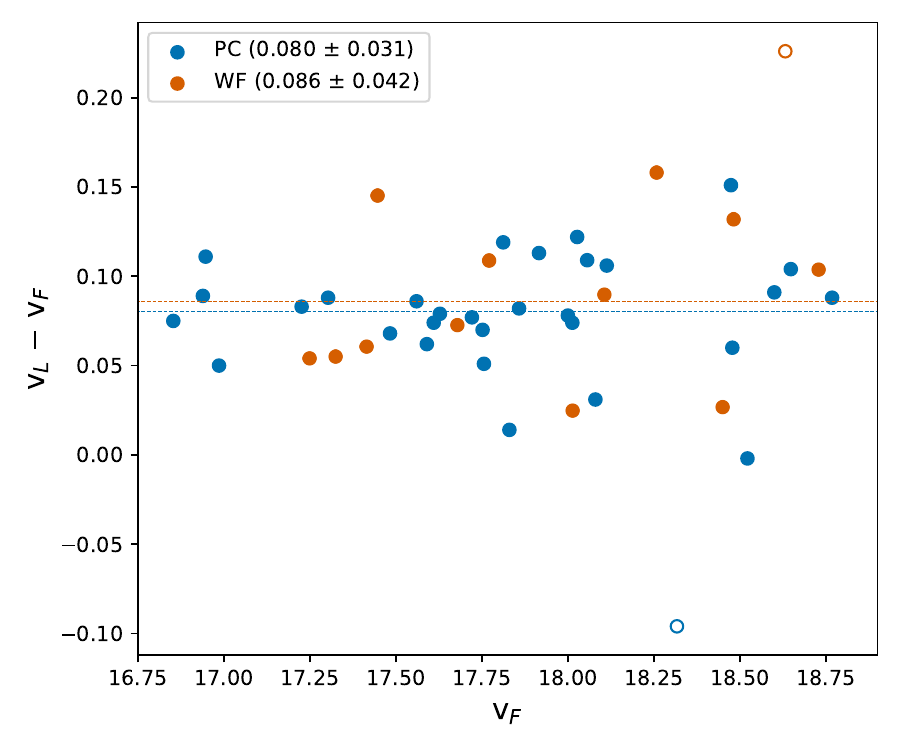}}
  \caption{Plot showing the v$_{555}$ magnitude differences between \citetalias{Lugger2007} (v$_L$) and \citetalias{Ferraro2009} (v$_F$) for BSSs located in the WFPC2/PC (Teal) and WFPC2/WF (Orange) frames. The outliers are shown as hollow markers. The mean $\pm$ standard deviation of the magnitude differences for the WFPC2/PC and WFPC2/WF frames are shown in the legend. Horizontal dashed lines in teal and orange show the mean magnitude difference for the BSSs in WFPC2/PC, and WFPC2/WF, frames respectively.} 
  \label{fig:vL_vF}
\end{figure}

We examined the v$_{555}$ magnitude differences between \citetalias{Lugger2007} and \citetalias{Ferraro2009} for BSSs located in the WFPC2/PC and WFPC2/WF frames (see Figure~\ref{fig:vL_vF}). Of the 31 BSSs in the WFPC2/PC frame, we found that the magnitude difference is $0.080 \pm 0.031$ after excluding one outlier. This outlier (B12) has a bright neighboring star that may affect its photometry (see Figure \ref{fig:findingcharts}). For the 13 BSSs in the WFPC2/WF frame, the magnitude difference is $0.086 \pm 0.042$ after excluding one outlier (B42), which has a neighboring faint star that might have affected its photometry, though the cause of the discrepancy remains unclear (see Figure \ref{fig:findingcharts}). The standard deviation of the magnitude differences in the WFPC2/PC frame (0.031) is comparable to the quadrature sum of the observed uncertainties in both datasets ($\approx 0.030$), suggesting that the differences between these photometric reductions are consistent with expected observational errors (these errors are likely dominated by systematics in the photometry methodology, rather than shot noise, as it is the same dataset). However, in the WFPC2/WF frame, the standard deviation of magnitude differences is larger (0.042), likely due to the larger pixel scale of the WFPC2/WF (0.10$^{\prime\prime}$) compared to the PC (0.046$^{\prime\prime}$) and the poorer quality of the drizzled WF images.

We also examined the v$_{555}$ magnitude differences between \citetalias{Lugger2007} and \citetalias{Ferraro2009} for SGB stars located in the WFPC2/PC and WFPC2/WF frames. After excluding five outliers showing large magnitude differences, we compared 110 SGB stars in the WFPC2/PC frame and 130 SGB stars in the WFPC2/WF frame.
The v$_{555}$ magnitude differences for SGB stars are found to be $0.108 \pm 0.045$ and $0.109 \pm 0.061$ in the WFPC2/PC and WFPC2/WF frames, respectively.
The standard deviation of magnitude differences is larger for SGB stars in the WFPC2/WF frame (0.061) compared to the WFPC2/PC frame (0.045), likely due to  larger pixel scale of the WF images. These values are larger than those measured for BSSs, likely because SGB stars are on average slightly fainter than BSSs in the CMD.

\subsection{HUGS Data} \label{sec:hugs_m30}
We cross-matched the 45 BSSs from \citetalias{Ferraro2009} to the HUGS NGC~7099 dataset without applying any data filtering (see \S~\ref{sec:hugs_filtering}). We successfully identified 43 matches in the HUGS catalog. The two unmatched BSSs, B12 and B25, are located in a region where the HUGS ACS/WFC photometry contains an unresolved saturation streak, preventing reliable measurements (see Figure \ref{fig:findingcharts}).

\begin{figure*}
\centering
\includegraphics[width=17cm]{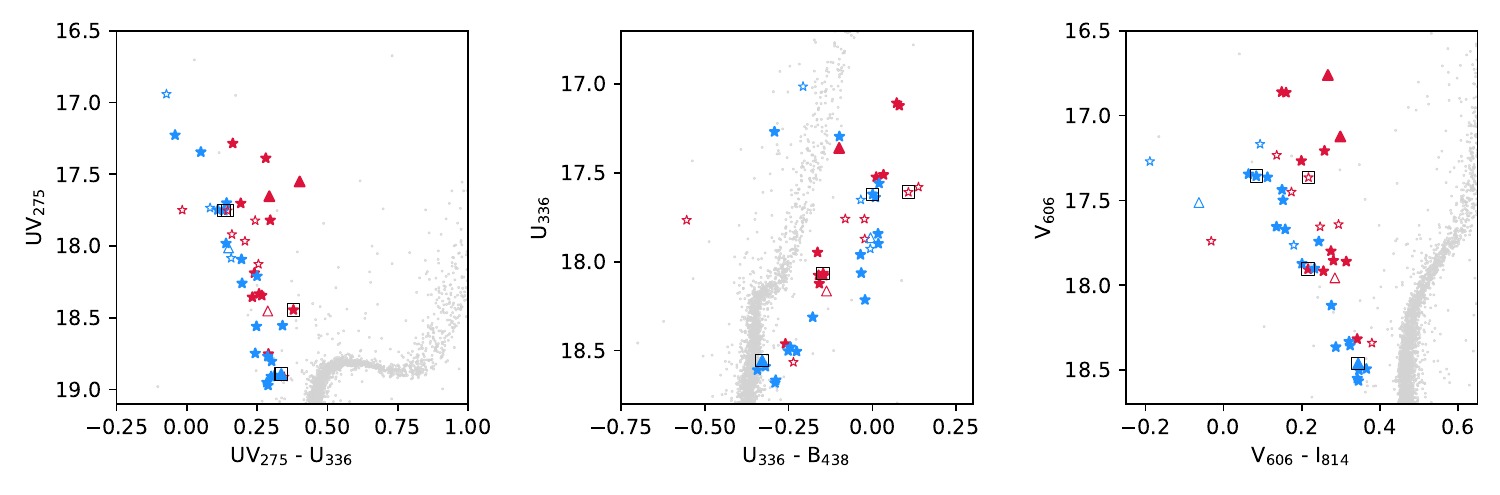}
\caption{HUGS CMDs of NGC~7099 (M30) showing the 43 BSSs from \citetalias{Ferraro2009} overlaid on the full unfiltered catalog. The BSSs are color-coded as blue and red based on the classification by \citetalias{Ferraro2009}. Marker shapes indicate data quality and membership: solid stars denote BSSs with membership probability $>90\%$ and good photometry in all filters; hollow stars indicate membership probability $>90\%$ but with at least one \texttt{SHARP} = $-1$ measurement; solid triangles mark membership probability $<90\%$ with good photometry; and hollow triangles represent membership probability $<90\%$ with at least one \texttt{SHARP} = $-1$. Markers surrounded with squares are W~UMa-type stars, taken from \citet{Clement2001}. }
\label{fig:m30_hugs_cmd}
\end{figure*}

Figure~\ref{fig:m30_hugs_cmd} shows the positions of these 43 BSSs in three different CMDs: UV$_{275}$--U$_{336}$ and U$_{336}$--B$_{438}$ from WFC3/UVIS, and V$_{606}$--I$_{814}$ from ACS/WFC.  We use different marker shapes to distinguish between stars with high cluster membership probability ($>90\%$) and low membership probability, as well as between stars with good photometry and those with \texttt{SHARP} = $-1$ in at least one filter, which indicates poor photometry.

The BSSs occupy the expected BSS region in both the UV$_{275}$--U$_{336}$ and V$_{606}$--I$_{814}$ CMDs. However, in the U$_{336}$--B$_{438}$ CMD, the presence of the Balmer jump causes the BSSs to become misaligned from their typical positions, appearing to the {\it red} of the subgiant branch. A similar inversion was visible in the $U-B$ CMD of  \citet{Guhathakurta98} (their Fig. 6), but was attributed by them to the red leak in the WFPC2 F336W filter, which is not a significant issue in the WFC3 F336W filter \citep{Brown08_redleak}. We attribute this inversion to two effects. First, the $U_{336}-B_{438}$ color is exquisitely aligned to demonstrate the effect of the Balmer jump at 3700-3800 \AA.  The Balmer jump is extremely prominent for F stars at around 7000 K, and highly dependent on surface gravity. On the other hand, the giant and subgiant branches become bluer as metallicity decreases. Thus, M30, as one of the very lowest metallicity clusters in the Galaxy ([Fe/H]=-2.27), has one of the bluest giant branches among globular clusters, allowing the BSSs to be redder than the giant branch in this color due to their Balmer jump. 

Using the method described in \S~\ref{sec:diptest_simulation}, we estimated the observational uncertainty for the unfiltered HUGS V$_{606}$--I$_{814}$ CMD of NGC~7099 to be $\sigma = 0.008$. This is significantly lower than the uncertainties derived for the \citetalias{Ferraro2009} ($\sigma = 0.020$) and \citetalias{Lugger2007} ($\sigma = 0.023$) optical CMDs. However, we expect that W UMa stars will experience substantial variability in this, and all, CMDs, due to their amplitude of variation of 0.27-0.75 magnitudes \citep{Clement2001}.

\section{Discussion} \label{sec:discussion}

Our systematic analysis of 56 HUGS globular clusters reveals no compelling statistical evidence for double BSS sequences. The Hartigan Dip Test \citep{Hartigan1985} applied to the BSS distance distributions in both UV$_{275}$--U$_{336}$ and V$_{606}$--I$_{814}$ CMDs yielded no $p$-values below 0.15. These high $p$-values are inconsistent with the presence of strongly bimodal distributions, even in clusters where such features have been previously reported in other data (see Table~\ref{tab:diptest_hugs}, Figures~\ref{fig:m30_hugs_diptest} and \ref{fig:diptest_hugs_4gc}). More importantly, our model comparison using $\mathrm{AIC_c}$ consistently favors a skewed unimodal Gaussian model over a mixture of two unskewed Gaussians across the vast majority of clusters. Only 17 out of 112 cases (6 in UV$_{275}$--U$_{336}$ and 11 in V$_{606}$--I$_{814}$ CMDs) show $\mathrm{ER} < 1$ favoring the mixture model. Even in these 17 cases, the evidence is generally weak and often associated with a (likely spurious) secondary peak comprising only $\sim$5--10 BSSs, inconsistent with the expected double BSS sequence morphology of similar numbers of BSSs in each sequence. Critically, among the five HUGS clusters in which double BSS sequences have been previously claimed (NGC~362, NGC~1261, NGC~6752, NGC~7078, and NGC~7099), only NGC~1261 (V$_{606}$--I$_{814}$ CMD) and NGC~7078 (UV$_{275}$--U$_{336}$ CMD) show ER $< 1$. However, even in these cases, the BSS distributions do not match the suggested morphology of bifurcated BSS sequences, and the evidence for bifurcation is not strong (ER of 0.4 and 0.2, respectively). The remaining three clusters, as well as NGC~1261 (UV$_{275}$--U$_{336}$ CMD) and NGC~7078 (V$_{606}$--I$_{814}$ CMD), exhibit ER values ranging from 1.2 to 276, corresponding to marginal to very strong evidence favoring the skewed unimodal model (see Table~\ref{tab:diptest_hugs}).

Among the globular clusters in which a double sequence has been reported, NGC~7099 (M30) represents the prototypical case \citepalias{Ferraro2009}. Our analysis using three independent \textit{HST} datasets provides an opportunity to assess the robustness of this detection across different instruments and reduction methods. The Hartigan Dip Test applied to the \citetalias{Ferraro2009} photometry, calculating a BSS distance distribution from a 1 Gyr reference isochrone, yields a $p$-value of $4 \times 10^{-3}$. Using the BSS distance distribution calculated from a straight line fitted to the blue BSS sequence (as done by \citetalias{Ferraro2009}), we obtained a $p$-value of $2 \times 10^{-3}$. Both these values differ significantly from their originally reported value of $\approx 1 \times 10^{-5}$, though F.~R. Ferraro indeed finds a similar $p$-value of $1.5\times10^{-3}$ (priv. comm.).  Since a double BSS sequence was not theoretically predicted before \citetalias{Ferraro2009}, it is quite possible that such a low p-value is a coincidence, not unusual when searching large parameter spaces for anomalies; i.e., the look-elsewhere effect  \citep[e.g.][]{BayerSeljak20}.

The independent \textit{HST}/WFPC2 reduction by \citetalias{Lugger2007}, using the same raw data but different reduction techniques, yields a dip test $p$-value of 0.162, indicating no significant evidence for bimodality. The comparison of BSS v$_{555}$ magnitudes between the two reductions reveals systematic differences with offsets of $0.080 \pm 0.031$~mag in the PC frame and $0.086 \pm 0.042$~mag in the WF frames.  The average offset is not important, but the standard deviation is interesting.  While the standard deviation in PC frame differences are consistent with expected observed uncertainties (quadrature sum $\approx 0.030$~mag), the larger scatter in the WF frames (0.042~mag) may reflect additional systematic effects in the photometric reductions. Our $\mathrm{AIC_c}$ model comparison for both the \citetalias{Ferraro2009} and \citetalias{Lugger2007} datasets consistently favors a skewed unimodal model, with ER values of $\approx$1.2 and $\approx$8, respectively, corresponding to marginal and positive evidence favoring the skewed unimodal Gaussian model over a mixture of two unskewed Gaussians.

We also examined another independent photometric analysis of the M30 \textit{HST}/WFPC2 data by \citet{Pietrukowicz2004}. Their CMD (their Figure~9) does not show a clear bifurcation in the BSS sequence, consistent with our findings. The double BSS sequence has also not been observed in the FUV--NUV (F150LP--F300W) CMD of M30 \citep{Mansfield2022}. Additionally, when we examined the BSSs reported by \citetalias{Ferraro2009} in the full unfiltered HUGS M30 catalog (Figure~\ref{fig:m30_hugs_cmd}), we did not observe a clear bifurcation in either the UV$_{275}$--U$_{336}$ or V$_{606}$--I$_{814}$ CMD. The HUGS V$_{606}$--I$_{814}$ CMD exhibits significantly lower observed uncertainty ($\sigma = 0.008$) compared to either \citetalias{Ferraro2009} ($\sigma = 0.020$) or \citetalias{Lugger2007} ($\sigma = 0.023$). This consistency across instruments (WFPC2 vs. WFC3/ACS), filters (v$_{555}$--i$_{814}$ vs. V$_{606}$--I$_{814}$, and FUV--NUV), and reduction methods suggests that the BSS distribution in M30 is better characterized as unimodal and skewed rather than clearly bifurcated.

We note that the statistical evidence reported for bifurcation in other globular clusters varies considerably, with different studies applying different statistical tests to the distribution of BSS distances from a best-fit line to the blue BSS sequence. This methodological heterogeneity can make direct comparisons challenging and raises questions about the robustness and reproducibility of these detections.

\citet{Dalessandro2013} reported a bimodal BSS sequence in NGC~362 with a confidence level of 3$\sigma$, using Gaussian mixture modeling to compare unimodal versus bimodal fits \citep{Muratov2010}. \citet{Simunovic2014} identified bifurcation in NGC~1261 by applying an Epanechnikov kernel density estimate, which revealed two peaks in the BSS distribution. Notably, their analysis also identified a third population of extremely blue BSSs (eB-BSSs) lying bluer than the blue BSS sequence. Interestingly, unlike other studies \citep{Ferraro2009, Dalessandro2013, Beccari2019, Cohn2021}, they found the blue BSSs to be more centrally concentrated than the red BSSs. 

\citet{Beccari2019} reported a double BSS sequence in NGC~7078 (M15) in UV CMDs, marking the first such detection outside optical wavelengths. They employed both the Hartigan Dip Test ($p$-value $\approx 0.015$) and Gaussian mixture modeling \citep{Muratov2010}, with both methods favoring a bimodal over a unimodal distribution. However, we note that their dip test $p$-value of 0.015, while formally significant at the $p < 0.05$ level, is substantially less significant than the $p \approx 1 \times 10^{-5}$ originally reported by \citetalias{Ferraro2009} for M30. To our knowledge, differential reddening has not been accounted for in the study of \citet{Beccari2019}. 
\citet{Cohn2021} reported a double BSS sequence in NGC~6752, though their study does not provide a specific statistical test or significance level for the bifurcation, instead using 
visual inspection of the CMD. \citet{Cadelano2022} identified a double BSS sequence in NGC~6256, using both the AIC and BIC to demonstrate that the BSS distribution is better reproduced by a sum of two Gaussians than by a single Gaussian, with a reported probability exceeding 99.5\% favoring bimodality. They also found no statistically significant difference between the cumulative radial distributions of the blue and red sequences, suggesting similar spatial distributions for both populations.

NGC~2173 is a young Large Magellanic Cloud cluster which shows no evidence of core collapse, and has a low central stellar number density. A double BSS sequence was first claimed by \citet{Li2018a, Li2018b} in this cluster, although subsequent analyses suggested that the bifurcation was likely due to field contamination \citep{Dalessandro2019a, Dalessandro2019b}. A recent study by \citet{Wang2025} confirmed that the BSS distribution in this cluster is indeed bifurcated. However, they calculate that the apparent bifurcation is consistent with apparent bifurcations found in 2\% of  simulations without real bifurcation. Combining this with the 20 young Magellanic Cloud star clusters as the number of trials (i.e., 5\% of such clusters appear to show this feature), this suggests that the apparent bifurcation may simply be a statistical coincidence.

The variety of statistical approaches employed across these studies, including the Hartigan Dip Test, Gaussian mixture modeling, kernel density estimation, AIC, and BIC, makes it difficult to assess whether the reported bifurcations represent a uniform physical phenomenon or artifacts of different analysis techniques and selection criteria. Furthermore, the reported significance levels vary widely, from marginal detections ($p \sim 0.01$--$0.05$) to highly significant ones ($p \sim 10^{-5}$).

A key hypothesis we tested in our study is that the BSS distance distributions can be better modeled by a skewed unimodal Gaussian than by a mixture of two unskewed Gaussians, whereas previous comparisons have focused exclusively on unimodal versus bimodal Gaussian models. The consistent preference for skewed unimodal models across the HUGS clusters suggests that the BSS distributions are naturally asymmetric rather than bifurcated. This asymmetry likely reflects the continuous nature of BSS formation and evolution rather than the coexistence of two distinct populations formed through different mechanisms. The skewed distribution may be interpreted as follows: BSSs are formed through both collisional and mass-transfer mechanisms, with many of them forming along a relatively narrow sequence that aligns with a young isochrone (corresponding to recent mass gain). As BSSs age, they evolve to higher luminosities and redder colors, creating an extended tail in the distribution towards the red (see Figure 1 of \citet{Sills2013}). This evolutionary sequence, from a concentrated peak of young BSSs to a more dispersed population of older, evolved BSSs, naturally produces a skewed unimodal distribution rather than two separate peaks. 

Variability of variable stars presents an additional complication for interpreting BSS sequences. As shown in Figures~\ref{fig:m30_hugs_diptest}, \ref{fig:m30_wfpc2_diptest}, and \ref{fig:diptest_hugs_4gc}, both W~UMa-type eclipsing contact (or semi-contact) binaries and SX~Phe pulsating variables can exist in the BSS sequence, with amplitudes of variation typically exceeding the gap between putative BSS sequences. Interestingly, W~UMa systems do contaminate the blue BSS region; 2 in M30 in Figure~1 of \citet{Ferraro2009}, while a 3rd appears to lie nearer the bluer BSSs in the HUGS M30 CMDs (Fig. \ref{fig:m30_hugs_cmd}). Figure~7 of \citet{Dalessandro2013} also shows a W UMa on the blue sequence of NGC 362. This is contrary to the hypothesis that blue and red BSS sequences correspond to collisional and mass-transfer formation channels, respectively, as W UMas are one class of mass transfer binary. Furthermore, theoretical modeling by \citet{Jiang2017} among others demonstrated that binary evolution can contribute to the formation of BSSs across the entire BSS color range in M30, challenging the interpretation that the blue sequence exclusively comprises collisional products. 
Collectively, these findings argue against the interpretation of bifurcated BSS sequences as clear signatures of two distinct formation mechanisms, and instead favor a picture in which both formation channels contribute to a continuous, asymmetric distribution shaped primarily by evolutionary effects.

\section{Conclusion} \label{sec:conclusion}

We have conducted a systematic search for double BSS sequences across 56 globular clusters in the HUGS survey and performed a detailed re-examination of NGC~7099 (M30), the prototypical case for this phenomenon. Our main findings are:

\begin{enumerate}

\item Hartigan Dip Test analyses of BSS distance distributions (from a young isochrone in the CMD) in both UV$_{275}$--U$_{336}$ and V$_{606}$--I$_{814}$ CMDs yield no significant evidence for bimodality across the HUGS sample, with no $p$-value below 0.15 in any cluster.

\item Akaike model comparison favors a skewed unimodal Gaussian model over a mixture of two unskewed Gaussians in 94 out of 112 cases (84\%). Critically, among the five HUGS clusters with previously claimed double sequences, only NGC~1261 (V$_{606}$--I$_{814}$ CMD) and NGC~7078 (UV$_{275}$--U$_{336}$ CMD) show ER $< 1$, weakly favoring the mixture model. However, even in these cases, the BSS distributions do not match the suggested morphology of bifurcated BSS sequences. All remaining cases exhibit ER values ranging from 1.2 to 276, corresponding to marginal to very strong evidence favoring the skewed unimodal model.

\item The BSS distributions in M30 were analysed from three reductions of \textit{HST} data (two WFPC2 reductions and the HUGS reduction of ACS and WFC3 data). The highly significant bimodality (Hartigan Dip Test $p \approx 1 \times 10^{-5}$) originally reported by \citetalias{Ferraro2009} could not be reproduced at that level of significance in our reanalysis (we find $p \approx 4 \times 10^{-3}$, or $p \approx 2 \times 10^{-3}$, depending upon the comparison line). The HUGS data, with smaller observational uncertainties, do not show evidence for such a bimodality. All BSS distributions from the three \textit{HST} reductions favor a skewed unimodal model over a mixture of two unskewed Gaussians, with ER values ranging from 1.2 to 94.

\end{enumerate}

Our results indicate that the observational evidence for double BSS sequences, while visually compelling in some cases, does not withstand rigorous statistical scrutiny across homogeneous datasets. The consistent preference for skewed unimodal distributions across diverse clusters, photometric systems, and reduction methods suggests that BSS populations follow a continuous distribution of properties rather than segregating into two distinct populations. This finding is consistent with 
theoretical predictions that both formation channels can produce BSSs spanning a continuous range of masses and evolutionary states \citep{Jiang2017}. We emphasize that the absence of clearly bifurcated sequences does not necessarily imply that only one formation mechanism operates in globular clusters. Rather, it suggests that the products of mass transfer and collisional formation may not be as cleanly separated in the CMD as previously thought.

\begin{acknowledgments}

We appreciate frank conversations with Francesco Ferraro regarding this work, and appreciate the use of his photometric data of \textit{HST}/WFPC2 for NGC~7099 (M30).

We also thank Natasha Ivanova for valuable discussions on the applicability of the finite-sample corrected AIC ($\mathrm{AIC_c}$), AIC, and BIC in various statistical contexts.

COH is supported by NSERC Discovery Grant RGPIN-2023-04264 and Alberta Innovates Advance Program \#242506334. 

This research has made use of the VizieR catalogue access tool, CDS, Strasbourg, France \citep{10.26093/cds/vizier}. The original description of the VizieR service was published in \citet{vizier2000}.

\end{acknowledgments}




\facilities{HST(ACS), HST(WFPC2), HST(WFC3)}

\software{Aladin sky atlas \citep{Aladin}, TOPCAT \citep{Topcat}, SAOImageDS9 \citep{DS9}}


\appendix

\section{Finding Charts for BSS\lowercase{s} with Problematic Photometry}

\begin{figure*}
\centering
\includegraphics[width=17cm]{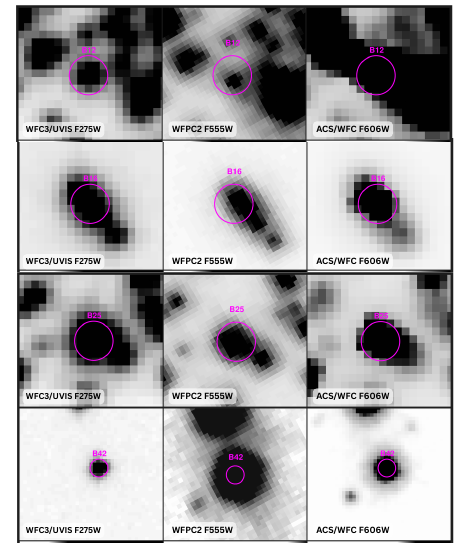}
\caption{Finding charts for BSSs with questionable photometry or non-detections. \textit{Left to right:} \textit{HST} WFC3/UVIS F275W, WFPC2 F555W, and ACS/WFC F606W images. B12 has a bright neighboring star affecting its photometry in both WFPC2 datasets (\S~\ref{sec:L07}). B16 appears extended and has zero HUGS membership probability (\S~\ref{sec:L07}). B25 falls on an unresolved saturation streak in the HUGS ACS/WFC data (\S~\ref{sec:hugs_m30}). B42 has a neighboring faint star to the lower-left in optical images that may affect its WFPC2/WF photometry (\S~\ref{sec:L07}). BSSs are marked with magenta circles of 0.10$^{\prime\prime}$ radius. B12, B16, and B25 are located within the WFPC2/PC frame, while B42 is in the WFPC2/WF frame.}
\label{fig:findingcharts}
\end{figure*}

Figure~\ref{fig:findingcharts} shows finding charts for the four BSSs that exhibit photometric issues or non-detections across the different datasets, as discussed in \S~\ref{sec:L07} and \S~\ref{sec:hugs_m30}.

\bibliography{references}{}
\bibliographystyle{aasjournalv7}

\end{document}